\documentclass[12pt]{article}

\usepackage{amsmath,amssymb,graphicx} %use drftcite in draftmode
\usepackage{pstricks}
\usepackage{cite}
\newcommand{\beq}{\begin{eqnarray}}% can be used as {equation} or  {eqnarray}
\newcommand{\eeq}{\end{eqnarray}}

\newcommand{\centeron}[2]{{\setbox0=\hbox{#1}\setbox1=\hbox{#2}\ifdim
\wd1>\wd0\kern.5\wd1\kern-.5\wd0\fi \copy0
\kern-.5\wd0\kern-.5\wd1\copy1\ifdim\wd0>\wd1
                                   \kern.5\wd0\kern-.5\wd1\fi}}
\newcommand{\ltap}{\>\centeron{\raise.35ex\hbox{$<$}}
                           {\lower.65ex\hbox{$\sim$}}\>}
\newcommand{\gtap}{\>\centeron{\raise.35ex\hbox{$>$}}
                           {\lower.65ex\hbox{$\sim$}}\>}

\newcommand{\lsim}{\mathrel{\ltap}}

\newcommand\ZZ{\hbox{\zfont Z\kern-.4emZ}}
\font\zfont = cmss10 %scaled \magstep1

\newcommand{\eref}[1]{eq.\ (\ref{e.#1})}
\newcommand{\erefn}[1]{ (\ref{e.#1})}

\newcommand{\cref}[1]{Chapter \ref{c.#1}}

\def\nn{\nonumber \\}
\newcommand{\nl}{& \nonumber \\ &}
\newcommand{\bnl}{\right . & \nonumber \\ & \left .}

\def\beq{\begin{equation}}
\def\eeq{\end{equation}}
\newcommand{\ba}{\begin{array}}
\newcommand{\ea}{\end{array}}
\newcommand{\bea}{\begin{eqnarray}}
\newcommand{\eea}{\end{eqnarray} }
\newcommand{\bal}{\begin{align}}
\newcommand{\eal}{\end{align}}
\def\bi{\begin{itemize}}
\def\ei{\end{itemize}}
\def\ben{\begin{enumerate}}
\def\een{\end{enumerate}}
\def\beq{\begin{equation}}
\def\eeq{\end{equation}}
\def\bc{\begin{center}}
\def\ec{\end{center}}
\def\bt{\begin{table}}
\def\et{\end{table}}
\def\btb{\begin{tabular}}
\def\etb{\end{tabular}}
\newcommand{\bvec}{\left ( \ba{c}}
\newcommand{\evec}{\ea \right )}

\def\cl{{\mathcal L}}

\def\co{{\mathcal O}}

\def\tev{\, {\rm TeV}}

\def\mass2{mass${}^2$}

\def\ra{\rangle}
\def\la{\langle}

\def\pa{\partial}

\newcommand{\tr}{\mathrm T \mathrm r}

\def\rt{\sqrt{2}}

\newcommand{\ha}{{\hat a}}

\newcommand{\ti}{\tilde}
\def\hc{{\rm h.c.}}

\def\ov{\overline}

\def\eps{\epsilon}

\begin{document}
\begin{titlepage}
%\begin{flushright}
%{\tt hep-ph/yymmnn}
%\end{flushright}

\vskip1.5cm
\begin{center}
{\huge \bf The Flavor of the Composite Pseudo-Goldstone Higgs}
\vspace*{0.1cm}
\end{center}
\vskip0.2cm

\begin{center}
{\bf Csaba Cs\'aki$^{a,b}$, Adam Falkowski$^{c,d}$, and Andreas
Weiler$^{a,b}$}

\end{center}
\vskip 8pt

\begin{center}
$^{a}$ {\it Institute for High Energy Phenomenology\\
Newman Laboratory of Elementary Particle Physics\\
Cornell University, Ithaca, NY 14853, USA } \\
\vspace*{0.3cm}
$^b$ {\it Kavli Institute for Theoretical Physics \\
University of California, Santa Barbara, CA 93106, USA} \\
\vspace*{0.3cm}
$^c$ { \it CERN Theory Division, CH-1211 Geneva 23, Switzerland} \\
\vspace*{0.3cm}
$^d$ {\it Institute of Theoretical Physics, Warsaw University,
           \\ Ho\.za 69, 00-681 Warsaw, Poland }

\vspace*{0.1cm}

{\tt  csaki@lepp.cornell.edu, adam.falkowski@cern.ch,
weiler@lepp.cornell.edu}
\end{center}

\vglue 0.3truecm

\begin{abstract}
\vskip 3pt \noindent We study the flavor structure of  5D warped
models that provide a dual description of a composite
pseudo-Goldstone Higgs. We first carefully re-examine the flavor
constraints on the mass scale of new physics in the standard
Randall-Sundrum-type scenarios, and find that the KK gluon mass
should generically be heavier than about 21 TeV. We then compare the
flavor structure of the composite Higgs models to those in the RS
model. We find new contributions to flavor violation, which while
still are suppressed by the RS-GIM mechanism, will enhance the
amplitudes of flavor violations. In particular, there is a kinetic
mixing term among the SM fields which (although parametrically not
enhanced) will make the flavor bounds even more stringent than in
RS. This together with the fact that in the pseudo-Goldstone scenario Yukawa couplings are set by a gauge coupling implies the KK gluon mass to be at least about 33 TeV. For
both the RS and the composite Higgs models the flavor bounds could
be stronger or weaker depending on the assumption on the value of
the gluon boundary kinetic term. These strong bounds seem to imply
that the fully anarchic approach to flavor in warped extra
dimensions is implausible, and there have to be at least some
partial flavor symmetries appearing that eliminate part of the
sources for flavor violation. We also present complete expressions
for the radiatively generated Higgs potential of various 5D
implementations of the composite Higgs model, and comment on the
$1-5$ percent level tuning needed in the top sector to achieve a
phenomenologically acceptable vacuum state.

\end{abstract}

\end{titlepage}

%\newpage

%\renewcommand{\thefootnote}{(\arabic{footnote})}

%%%%%%%%%%%%%%%%%%%%%%%%%%%%%%%%%%%%%%%%%%%%%%%%%%%%%%%%%%%%%%%%%%%%%%
%%%%%%%%%%%%%%%%%%%%%%%%%%%%%%%%%%%%%%%%%%%%%%%%%%%%%%%%%%%%%%%%%%%%%%%
\section{Introduction}
\label{sec:Intro} \setcounter{equation}{0} \setcounter{footnote}{0}
%%%%%%%%%%%%%%%%%%%%%%%%%%%%%%%%%%%%%%%%%%%%%%%%%%%%%%%%%%%%%%%%%%%%%%
%%%%%%%%%%%%%%%%%%%%%%%%%%%%%%%%%%%%%%%%%%%%%%%%%%%%%%%%%%%%%%%%%%%%%%

Warped extra dimensions models were introduced by Randall and
Sundrum  (RS) \cite{RS} as an attempt to solve the hierarchy problem
by making use of the warp factor to lower the natural scale of
particle masses. In the original model, all SM fields were localized
on the TeV brane. By the AdS/CFT duality, this corresponds to a
situation where a strongly coupled  4D conformal field theory
spontaneously breaks conformality at the TeV scale, creates a mass
gap (confines), and produces the SM fields as approximately massless
composites. One consequence of this scenario is the CFT cannot be
flavor invariant, since it is supposed to produce the Yukawa
couplings among the SM fields. In such a case,  however, the CFT is
also expected to generate higher-dimensional flavor violating
operators with only a TeV suppression, which would be disastrous
from the phenomenological point of view. Phrased in the 5D language
this question is why generic TeV localized four-fermion operators
are suppressed by some high scale, rather than the local cut-off
scale which is a few TeV.

This severe flavor problem can be avoided by instead considering
setups with the SM gauge fields and fermions in the bulk, and only
the Higgs sharply localized on the TeV brane \cite{P,GN,GP}.
In this case the SM fermions can be thought of as mixtures of elementary and composite fermions.
The amount of mixing is determined by the profile of the 5D wave
function of the fermions: the more peaked they are close to the
Planck brane, the more elementary are these fields. This sheds some
light on  the flavor puzzle as well: Although the Yukawa couplings
generated by the CFT are indeed  all ${\cal O}(1)$ and non-diagonal,
the fermion masses and the CKM angles depend also on the amount of
mixing of the elementary fermions with the CFT that is assumed to be
small for the first two generations~\cite{GN,GP,H}. Most
importantly, this implies that flavor violation in the SM is also
suppressed by the same mixing factors - the fact that goes under the
name of the RS-GIM mechanism \cite{GP,APS,APPP}, see also \cite{Burdman:2002gr,DIU}.
RS-GIM is successful in suppressing most of the dangerous flavor-changing neutral
currents (FCNC) \cite{APS}, although it is not enough to
sufficiently suppress new physics contributions to CP violation in
the kaon sector \cite{UTfit,A}.

Although the RS set-up explains the origin of the large
$M_{PL}/\tev$ hierarchy, there still remains the little hierarchy
problem, which amounts to the question   why the Higgs boson is much
lighter than a few TeV. In RS, the Higgs is realized as a scalar
field localized on the TeV brane, of which the 4D dual
interpretation is that Higgs a composite state of the CFT. In that
case, its mass would naturally be at the CFT scale of a few TeV,
which would effectively produce a Higgsless model of the sort
considered in~\cite{higgsless}. One way to obtain a light composite
Higgs is by making it a pseudo-Goldstone boson (pGB) of a global
symmetry~\cite{GK}, similarly as in little Higgs models~\cite{lh}.
The 5D holographic version of this scenario are the models of gauge-Higgs
unification (GHU)~\cite{M}, where the Higgs boson is identified with
the fifth component of a bulk gauge field ($A_5$) \cite{CNP}.
The Higgs potential is radiatively generated (with the largest contributions
due to the top and gauge multiplets) and fully calculable. The most
promising scenario of this kind is based on the SO(5) gauge group in
the bulk, broken spontaneously to SO(4) on the TeV brane \cite{ACP},
and with an implementation of a discrete parity symmetry to control
corrections to the $Z b\bar b$ vertex \cite{ACDP}. This is the
minimal scenario that passes the stringent electroweak precision
tests.

Thus, GHU provides us with  the Higgs sector that allows one  to
address both the large and the little hierarchy problem. It is a
natural question to ask how this dynamical realization of the Higgs
boson  affects the flavor structure and the RS-GIM mechanism. This
is the subject of this paper.

Perhaps not surprisingly, the flavor structure of GHU turns out to
be quite similar to that in RS. However, there are some important
differences, that affect the flavor bounds. The gauge symmetry must
be larger than in RS (to include broken generators that  give rise
to the pseudo-Goldstone Higgs degrees of freedom), which results in
a different embedding of the SM fermion into the bulk
representation. In particular, one SM fermion must be embedded in
several bulk multiplets. This induces a new effect not present in
RS, namely that, in the original flavor basis, the various
fermionic generations are mixed via kinetic terms. This kinetic
mixing is a new source of flavor violation in GHU which is always
non-zero (as long as the CKM mixing is reproduced). Even though the
kinetic mixing respects the RS-GIM mechanism, it results in an
enhancement of the flavor violating operators. As a consequence, the
bounds on the scale of the extra dimensions from flavor physics
turn out to be more stringent compared to the RS
case.

This paper is organized as follows: in Section~2 we review the
flavor structure and the RS-GIM mechanism of ordinary RS models with
gauge and fermions in the bulk and the Higgs on the TeV brane. We
reevaluate the bounds on the mass of the KK gluon and find it has to
be $\geq 22$ TeV. In Section~3 we review the basics of the GHU
models and then define three different realizations based on the
SO(5) gauge symmetry in the bulk. Before investigating the flavor
structure,  we study the constraints on the parameter space of these
models imposed by the requirement of correct electroweak symmetry
breaking. In Section~4 we calculate the Higgs potential for the
three SO(5) models. We re-emphasize that one of the parameters of
the top sector has to be tuned at the 1-5 percent level to end up
with a realistic scenario, which is a new guise of the little
hierarchy problem specified to GHU models. These relations among the
parameters of the model are then used as an input in our studies of
the flavor bounds. Section~5 contains our main results. There we
present the flavor structure of the SO(5) models under
investigation, and pinpoint new sources of flavor violation. We
estimate the magnitude of  flavor and CP violation induced by the KK
gluon exchange and illustrate our analytical estimates by numerical
scans over the parameter space allowed by electroweak symmetry
breaking. We conclude with some remarks on future directions of GHU
in light of our findings.

%%%%%%%%%%%%%%%%%%%%%%%%%%%%%%%%%%%%%%%%%%%%%%%%%%%%%%%%%%%%%%%%%%%%%%%%%%
%%%%%%%%%%%%%%%%%%%%%%%%%%%%%%%%%%%%%%%%%%%%%%%%%%%%%%%%%%%%%%%%%%%%%%%%
\section{Flavor in RS}
\label{sec:frs} \setcounter{equation}{0} \setcounter{footnote}{0}
%%%%%%%%%%%%%%%%%%%%%%%%%%%%%%%%%%%%%%%%%%%%%%%%%%%%%%%%%%%%%%%%%%%%%%%%%%%%
%%%%%%%%%%%%%%%%%%%%%%%%%%%%%%%%%%%%%%%%%%%%%%%%%%%%%%%%%%%%%%%%%%%%%%%%%%%%%

The original motivation for considering warped extra dimension was
the solution to the hierarchy problem of the Higgs sector. It was
quickly realized that the set-up also has the potential to explain
simultaneously the SM flavor structure.
Starting with completely anarchic Yukawa coupling of the Higgs and the 5D
fermions, the large SM fermion mass hierarchies can be explained by
different localization the SM fermions in the extra dimension~\cite{GN,GP,Huber:2000ie},
implementing the split fermion scenario of~\cite{AS}.
Small mixing angles of the CKM matrix are a natural consequence of this scenario~\cite{H}.
Moreover, this way of generating flavor mass
hierarchies automatically implies a certain amount of suppression of
the dangerous flavor changing, which is referred to as the RS-GIM
mechanism.
In this section  we will review the
flavor bounds on the generic RS models with anarchic flavor
structure. This will provide us with a reference point for our study
of the flavor bounds in the GHU models.

We specify the background metric to be AdS$_5$ space. % though the results can be easily generalized to an arbitrary metric.
We parametrize the space-time by the conformal coordinates
\begin{equation}
ds^2=\left( \frac{R}{z}\right)^2 (dx_\mu dx_\nu \eta^{\mu\nu} -dz^2) \, ,
\end{equation}
where the AdS curvature is $R$,  and the coordinate $z$ of the extra dimension runs between $R<z<R'$,
$z=R$ corresponding to the UV (Planck) brane and $z=R'$ to the IR (TeV) brane.
$R'/R \sim 10^{16}$ sets the large hierarchy between the Planck and the TeV scale.

We consider here the standard RS scenario with custodial symmetry \cite{ADMS} (this is always what we mean when we refer to RS in the following).
The bulk gauge group  $SU(3)_C \times SU(2)_L\times SU(2)_R \times U(1)_X$ and the Higgs field  transforming as $(1,2,2)_0$ is localized
on the TeV brane. The fermionic content includes three copies of
$\Psi_q^i$, $i = 1\dots 3$, transforming as $(2,1)_{1/6}$, and 3
copies each of $\Psi_{u,d}^i$ in  $(1,2)_{1/6}$. Each of these
fields are 5D bulk Dirac spinors and have a bulk mass term which is
customarily parametrized by using the $c$-parameters,
\begin{equation}
\left( \frac{R}{z}\right)^4 \left[ \frac{c_q^i}{z} \bar{\Psi}_q^i
\Psi_q^i+ \frac{c_u^i}{z} \bar{\Psi}_u^i \Psi_u^i+\frac{c_d^i}{z}
\bar{\Psi}_d^i \Psi_d^i\right].
\end{equation}
From now on we drop the generation index $i$; all fermions should
always be understood as  three-vectors in the generation space. The
boundary conditions on the UV and the IR brane are chosen as \beq
\Psi_{q} =  \bvec q [+,+] \evec \qquad \Psi_{u} =  \bvec u^c [-,-]
\\
\ti d^c [+,-]    \evec
\qquad
\Psi_{d} =  \bvec
\ti u^c [+,-]
\\
d^c [-,-]   \evec \eeq where $[\pm]$ denotes the
right(left)-chirality of a bulk fermion vanishing on the brane. The
SM quark doublets are realized as zero-modes $q$, while the singlets
up and down-type quarks are  zero modes of  $u^c$ and $d^c$,
respectively. We write it as \beq q(x,z) \to \chi_q(z) q_L(x) \qquad
u^c(x,z) \to \psi_u(z) u_R(x) \qquad d^c(x,z) \to \psi_d(z) d_R(x)
\eeq where $\chi_q(z)$ and $\psi_{u,d}(z)$ are the zero mode
profiles that are obtained by solving the equations of motion. A
normalized left-handed zero-mode profile is given by
\begin{equation}
\label{e.zm}
\chi_c(z) = R'{}^{-1/2}  \left( \frac{z}{R}\right)^{2}  \left( \frac{z}{R'}\right)^{-c} f(c),
\qquad
\psi_c(z) = R'{}^{-1/2}  \left( \frac{z}{R}\right)^{2}  \left( \frac{z}{R'}\right)^{c} f(-c) \, .
\end{equation}
We have introduced the standard RS flavor function
$f(c)$, which is given by
\begin{equation}
\label{e.rsff}
f(c)=\frac{\sqrt{1-2 c}}{\left[1-(\frac{R'}{R})^{2c-1}\right]^{\frac{1}{2}}}.
\end{equation}
We also introduce 3$\times$3 diagonal matrices $f_c$ that are
constructed from $f(c_i)$ of three generations, for example   $f_q =
{\rm diag}(f(c_{q_1}),f(c_{q_2}),f(c_{q_3}))$. For the choice of the
$c$-parameters that reproduce the SM mass hierarchies the matrices
$f_{q,-u,-d}$ are exponentially hierarchical.

The masses of the zero modes come from the IR brane-localized Yukawa
interactions. After  the Higgs  field acquires a vev, the Yukawa
terms lead to IR localized mass  terms for the bulk fermions, \beq
\label{e.rsyuk} \cl_y = - {v \over \sqrt 2} (R^4/R'{}^3) \left ( \ov
\Psi_q \ti Y_u \Psi_u + \ov \Psi_q \ti Y_d \Psi_d \right ) + \hc
\eeq The brane Yukawa couplings $\ti Y_{u,d}$ are assumed to be
anarchic -- random matrices with elements ${\cal O}(1)$, no
hierarchy and ${\cal O}(1)$ determinant.

Inserting the zero mode profiles into the mass terms in \eref{rsyuk} we obtain the SM mass matrices
\bea
m_{u}^{SM} &=& \frac{v}{\sqrt{2}}f_q \ti Y_{u} f_{-{u}},
\nn
m_{d}^{SM} &=& \frac{v}{\sqrt{2}}f_q \ti Y_{d} f_{-{d}},
\label{RSmass}
\eea
From  this point on the  usual SM prescription applies.
We diagonalize the up and down mass matrices  by $m_{u,d}^{SM}=U_{L\ u,d} m_{u,d} U_{R\ u,d}^\dagger$,
where $U$'s are unitary and $m_{u,d}$ are diagonal, and we rotate the zero modes to the mass eigenstate basis,
for example $d_L(x) \to U_{L\ d} d_L(x)$.
The left rotations yield the CKM matrix, $V_{CKM}=U_{L\ u}^\dagger U_{L\ d}$.

Even though the Yukawa matrices are anarchical, the hierarchical
matrices $f_c$ introduce the hierarchy  into  the mass matrix
elements.
\begin{table}
\begin{center}
\begin{tabular}{@{}cc}
\hline\hline
& value at 3 TeV \\
\hline
$\bar m_u$  & $0.00075 \ldots 0.0015$ \\
$\bar m_c$  & $0.56 \pm 0.04 $ \\
$\bar m_t$  & $136.2 \pm 3.1$ \\
\hline
$\bar m_d$  & $0.002 \ldots 0.004$  \\
$\bar m_s$  & $0.047 \pm 0.012$ \\
$\bar m_b$  & $2.4 \pm 0.04$ \\
\hline
\end{tabular}
\end{center}
\caption {$\overline{MS}$ quark masses in GeV at 3 TeV. We have taken the ranges and low-energy values from PDGLive~\cite{Yao:2006px} and used
LO renormalization equations with the appropriate number of flavors for the rescaling.
At 30 TeV, the masses $\bar m_i$ are about $11 \%$ smaller.}
\label{tab:quarkTeV}
\end{table}
This will result in the hierarchy of the eigenvalues\footnote{Here
and in the following, the quark masses are understood as running
masses at the scale at which the extra dimension is integrated out.
We choose the scale of the extra dimension to be 3 TeV. In Table
\ref{tab:quarkTeV} we collect all input values at this scale.}: \beq
(m_{u,d})_{ii} \sim  {v \over \sqrt 2} Y_*  f_{q_i} f_{-u_i,d_i}
\eeq
where $Y_*$ is the typical amplitude of the entries in the
Yukawa matrices. One can also show that the diagonalization matrices
themselves are hierarchical~\cite{H}:
\begin{equation}
|U_{L \ ij}| \sim \frac{f_{q_i}}{f_{q_j}}, \ \ |U_{R \ ij}| \sim
\frac{f_{-{u,d}_i}}{f_{-{u,d}_j}}, \ \ {\rm} \ i\leq j.
\end{equation}
We also get that $|V_{CKM}|_{ij} \sim f_{q_i}/f_{q_j}$, thus the
hierarchy in the CKM matrix elements is purely set by the $c_q$ parameters.
From experiment we know that the hierarchy of the CKM matrix is of the form
\begin{equation}
V_{CKM} \sim \left( \begin{array}{ccc} 1-\frac{\lambda^2}{2} &
\lambda & \lambda^3 \\ \lambda & 1-\frac{\lambda^2}{2} & \lambda^2
\\ \lambda^3 & \lambda^2 & 1 \end{array} \right)
\end{equation}
where $\lambda \sim \sin \theta_{C} \sim 0.2$.
This fixes the hierarchy among the $f_{q_i}$'s to be \cite{H}
\begin{equation} f_{q_2}/f_{q_3} \sim \lambda^2, \qquad f_{q_1}/f_{q_3} \sim \lambda^3.
   \end{equation}
The values of $f_{-u,-d}$ are then fixed by requiring that the correct fermion mass hierarchy is reproduced, implying
the following relations (assuming $f_{-u_3}\sim {\cal O}(1)$):
\begin{equation} \label{equ:crhRS}
f_{-d_3}\sim \frac{m_b}{m_t},\quad
f_{-u_2} \sim \frac{m_c}{m_t} \frac{1}{\lambda^2},\quad
f_{-d_2} \sim \frac{m_s}{m_t} \frac{1}{\lambda^2},\quad
f_{-u_1} \sim \frac{m_u}{m_t} \frac{1}{\lambda^{3}},\quad
f_{-d_1} \sim \frac{m_d}{m_t}\frac{1} {\lambda^{3}}.
   \end{equation}
Thus, the RS set-up leads to a neat explanation of the SM flavor
structure. However, one potentially worrisome feature of
higher-dimensional models is the presence of  the new KK states,
whose masses are in the TeV range (as long as the hierarchy problem
is addressed). These new states generically have flavor
non-universal couplings and will contribute to flavor-changing
neutral currents.

The largest contribution to flavor changing neutral currents is
generated via the exchange of heavy  gauge bosons, in particular the strongest constraint arises from the exchange of the KK gluons.
In order to calculate the effective four-Fermi operators we first need to determine the couplings $g_x$ of the zero-modes to the KK gluons,
\begin{equation}
\label{e.kkgc} g_{L,u}^{ij} \bar{u}_{L}^i \gamma_\mu G^{\mu (1)} u_{L}^j
+g_{L,d}^{ij} \bar{d}_{L}^i \gamma_\mu  G^{\mu (1)} d_{L}^j
+(L\to R)
\end{equation}
Below we  discuss the contribution of the lightest KK gluon but,
as we show in Appendix \ref{a.kkgs}, it is possible to sum up the contribution of the entire gluon KK tower.
The profile of the first KK gluon
can be approximated by
$G^{(1)}(z)\simeq \frac{\sqrt{2}}{J_1(x_1) \sqrt{R}R'} z J_1(x_1 z/R')$
with $x_1$ being the first zero of the Bessel function, $J_0(x_1)=0$.
Using this and the zero mode profiles we can determine the couplings in \eref{kkgc}.
In the original flavor basis the couplings are diagonal and well approximated by
\begin{eqnarray}
g_{x} \approx  % g_s \sqrt{\log R'/R}
g_{s*} \left(
-\frac{1}{\log R'/R} +  f_x^2 \,\gamma(c_x)
 \right).
\end{eqnarray}
where $g_{s*}$ is the bulk $SU(3)$ gauge couplings, and
$\gamma(c) = \frac{\sqrt{2} x_1}{J_1(x_1)} \int_0^1 x^{1-2 c} J_1(x_1 x) dx \approx \frac{\sqrt{2} x_1}{J_1(x_1)}  \frac{0.7}{6-4 c}$.\footnote{The function $\gamma(c)$ is a correction to the approximation used in~\cite{APS} which can be sizable, e.g. $\gamma(-0.4)\approx 0.52,\gamma(0.5)\approx 1.16,\gamma(0.7)\approx 1.52$. In numerical calculations we always use the full overlap integral.}
The couplings would  be flavor universal if $f_x \sim 1_{3\times 3} $.
However, this not the case in the RS scenario where the $f_x$ are non-degenerate. This is the main source of flavor
violation in RS. Going to the mass eigenstate basis, we have to
rotate the couplings appropriately, \beq \label{e.rsg} g_{L,u,d} \to
U_{L\ u,d}^\dagger g_{q} U_{L\ u,d} \qquad g_{R,u,d} \to U_{R\
u,d}^\dagger g_{-u,-d} U_{R\ u,d} \eeq The rotation introduces non
diagonal couplings which lead to tree level contributions to $\Delta
F = 2$ processes. Nevertheless, the rotation matrices are
hierarchical, with the hierarchy set by the same $f_x$ that controls
the SM fermion hierarchies.
The off-diagonal KK gluon couplings are of order
\beq \label{e.rsge} (g_{L,q})_{ij} \sim   g_{s*}  f_{q_i}  f_{q_j} \quad
(g_{R,u})_{ij} \sim   g_{s*} f_{-u_i}  f_{-u_j}
\quad (g_{R,d})_{ij} \sim   g_{s*} f_{-d_i}  f_{-d_j}
\eeq
The off-diagonal couplings of the quark
doublets are suppressed by the ratios of the CKM matrix elements
(recall that $f_{q_1} \sim \lambda^3$, $f_{q_2} \sim \lambda^2$).
Similarly, the off-diagonal couplings of the singlet quarks are
suppressed by hierarchically small entries. This suppression is
called the RS-GIM mechanism. It is enough to suppress most of the
dangerous $\Delta F = 2$ operators, though not all, as we will see
in a moment.

Integrating out the KK gluon and applying appropriate Fierz identities we obtain the  effective Hamiltonian:
\begin{eqnarray}
\mathcal{H} &=&\frac{1}{M_G^2} \left[
\frac{1}{6} g_L^{ij} g_L^{kl}
(\bar{q}_L^{i \alpha} \gamma_\mu q_{L \alpha}^j)\ (  \bar{q}_L^{k \beta} \gamma^\mu q_{L \beta}^l)
- g_R^{ij} g_L^{kl} \left( (\bar{q}_R^{i \alpha} q_{L
\alpha}^k)\
(\bar{q}_L^{l \beta} q_{R \beta}^j)
-\frac{1}{3}  (\bar{q}_R^{i \alpha} q_{L
\beta}^l)\
(\bar{q}_L^{k \beta} q_{R \alpha}^j)\right)\right]\nonumber\\  &=& C^1(M_G) (\bar{q}_L^{i \alpha} \gamma_\mu q_{L \alpha}^j)\ (  \bar{q}_L^{k \beta} \gamma^\mu q_{L \beta}^l) +  C^4(M_G)    (\bar{q}_R^{i \alpha} q_{L
   \alpha}^k)\
   (\bar{q}_L^{l \beta} q_{R \beta}^j) +   C^5(M_G) (\bar{q}_R^{i \alpha} q_{L
   \beta}^l)\
   (\bar{q}_L^{k \beta} q_{R \alpha}^j)
\nonumber
\end{eqnarray}
where $\alpha,\beta$ are color indices. The Wilson coefficients of
these operators will directly correspond to the $C^{1,4,5}$ bounded
by the model independent constraints from $\Delta F =2$ processes by
the UTFit collaboration\footnote{We are grateful to Luca Silvestrini for discussions about the proper interpretation of the UTFit bounds.} in~\cite{UTfit}, see Table~\ref{tab:all}.
\renewcommand{\arraystretch}{1.5}
\begin{table}
\begin{center}
\begin{tabular}{@{}lll}
\hline\hline
Parameter & Limit on $\Lambda_F$ (TeV) & Suppression in RS (TeV)\\
\hline
Re$C_K^{1}$  & $1.0 \cdot 10^{3}$ & $\sim r /( \sqrt{6} \, |V_{td} V_{ts}| f_{q_3}^2) = 23 \cdot 10^{3}$ \\
Re$C_K^{4}$  & $12 \cdot 10^{3}$ &  $\sim r (v Y_*)/(\sqrt{2\, m_d m_s})=22 \cdot 10^{3} $ \\
Re$C_K^{5}$  & $10 \cdot 10^{3}$ &  $\sim r (v Y_*)/(\sqrt{6\, m_d m_s}) =38 \cdot 10^{3} $\\
\hline
Im$C_K^{1}$  & $15 \cdot 10^{3}$ & $\sim r/(\sqrt{6}\, |V_{td} V_{ts}| f_{q_3}^2) = 23 \cdot 10^{3}$ \\
Im$C_K^{4}$  & $160 \cdot 10^{3}$ &$\sim r (v Y_*)/(\sqrt{2\, m_d m_s})=22 \cdot 10^{3} $ \\
Im$C_K^{5}$  & $140 \cdot 10^{3}$ &$\sim r (v Y_*)/(\sqrt{6\, m_d m_s}) =38 \cdot 10^{3} $\\
\hline
\hline
$|C_{D}^{1}|$ & $1.2 \cdot 10^{3}$ & $\sim r/(\sqrt{6}\, |V_{ub} V_{cb}| f_{q_3}^2) = 25 \cdot 10^{3}$ \\
$|C_{D}^{4}|$ & $3.5 \cdot 10^{3}$ & $\sim r (v Y_*)/(\sqrt{2\, m_u m_c})=12 \cdot 10^{3} $ \\
$|C_{D}^{5}|$ & $1.4 \cdot 10^{3}$ & $\sim r (v Y_*)/(\sqrt{6\, m_u m_c}) =21 \cdot 10^{3} $\\
\hline
\hline
$|C_{B_d}^{1}|$  & $0.21 \cdot 10^{3}$& $\sim r/(\sqrt{6}\, |V_{tb} V_{td}| f_{q_3}^2) = 1.2 \cdot 10^{3}$ \\
$|C_{B_d}^{4}|$  & $1.7 \cdot 10^{3}$ & $\sim r (v Y_*)/(\sqrt{2\, m_b m_d})=3.1 \cdot 10^{3} $ \\
$|C_{B_d}^{5}|$  & $1.3 \cdot 10^{3}$ & $\sim r (v Y_*)/(\sqrt{6\, m_b m_d}) =5.4 \cdot 10^{3} $\\
\hline
\hline
$|C_{B_s}^{1}|$ & $30$  & $\sim r/(\sqrt{6}\, |V_{tb} V_{ts}| f_{q_3}^2) = 270$ \\
$|C_{B_s}^{4}|$  & $230$ &$\sim r (v Y_*)/(\sqrt{2\, m_b m_s})= 780$ \\
$|C_{B_s}^{5}|$  & $150$ &$\sim r (v Y_*)/(\sqrt{6\, m_b m_s}) = 1400$\\
\hline
\hline
\end{tabular}
\end{center}
\caption {Lower bounds on the NP flavor
scale $\Lambda_F$ for arbitrary NP flavor structure from~\cite{UTfit} and the effective suppression scale in RS for KK mass with $M_G=3$ TeV. Since the Wilson coefficients in~\cite{UTfit} are given at the scale $\Lambda_F$, we have corrected for the renormalization group scaling from $\Lambda_F$ to 3 TeV using the expressions in~\cite{Buras:2001ra} when necessary.
We have set $|Y_*| \sim 3$, $f_{q_3}=0.3$ and $r = M_G/g_{s*}$.
%Note, that for the $D-\bar D$ we compared the $RR$ coefficient $|\ti C_{D}^{1}|$ to the bound.
}
\label{tab:all}
\end{table}
\renewcommand{\arraystretch}{1}
Note, that the most strongly constrained quantity is the imaginary part of $C^4_K$ for the kaon system. Contributions to $\epsilon_K$ coming from $C^4_K$ are enhanced compared to the ones to $C^1_K$ with SM like chirality by
\begin{equation}
\sim \frac{3}{4} \left( \frac{m_K}{m_s(\mu_L) + m_d(\mu_L)} \right)^2 \eta_1^{-5}
\end{equation}
where the first factor $\approx 18$ is the chiral enhancement of the hadronic matrix element and $\eta_1^{-5}\approx 8$ is the relative RGE running~\cite{Buras:2001ra}.

We are ready to estimate the flavor bounds of the RS model.
Using the expressions for the orders of magnitudes for the rotation matrices $U$ we
approximately find for the Wilson coefficient at the TeV scale
\begin{equation}
C_{4 K}^{RS} \sim  \frac{g_{s*}^2} {M_G^2} f_{q_1} f_{q_2} f_{-d_1} f_{-d_2}
\sim
\frac{1}{M_G^2} \frac{g_{s*}^2}{Y_*^2}  \frac{2 m_d m_s}{v^2} .
\end{equation}
Above, $m_d, m_s$ are the down, strange masses at the  TeV scale,
see Table~\ref{tab:quarkTeV}. At tree level and in absence of
boundary kinetic terms, the bulk coupling $g_{s*}$ is connected the
strong coupling at the KK scale by $g_{s*} = g_s(M_G)
\log^{1/2}(R'/R) \sim 6$. If arbitrary large $Y_*$ was allowed the
bounds from $C_{4K}$ could be eliminated completely. However, if
$Y_*$ is too large one loses perturbative control over the theory.
One can estimate the upper bound on $Y_*$ using naive dimensional
analysis.
The proper brane localized Yukawa coupling $Y_{5D}$ is in our normalization $Y_{5D}= Y_* R'$.
One-loop corrections to the localized Yukawa coupling would be proportional to $Y_{5D} (Y_{5D} E)^2 / (16 \pi^2)$,
where the energy dependence is inferred from the dimension $-1$ of $Y_{5D}$.
In order for this to be smaller than the tree-level term we need to impose $(Y_{5D} E)^2/(16\pi^2) \ltap 1$.
We require that this bound is not violated until we reach the energies $N$ KK modes, $E = N_{KK} m_{KK}$.
Using $m_{KK}\sim 2/R'$ we find that  $Y_* \ltap (2\pi)/N_{KK}$. For the most conservative bound we set $N_{KK}=2$,
which imposes $Y_* \ltap 3$. Thus in our estimates we assume $|Y_*| \ltap 3$.
The suppression scale of the four-fermion operator is set by the lightest KK gluon mass $M_G$. The RS-GIM mechanism effectively
raises the suppression scale by the factor $v/\sqrt{m_d m_s} \sim
10^{4}$. However, that factor turns out to be an order of magnitude
too small   for a $\sim 3 \tev$ KK gluon. One can find that in order
for the suppression scale to match $1.6 \cdot 10^5$ TeV we need $M_G
\sim (22 \pm 6)$ TeV. Our estimate
is less optimistic than that encountered in the RS literature so
far~\cite{A,MS,FPRa}. The quoted error comes from the uncertainty in
the $m_d$ and $m_s$ masses. Including the contributions of the full
KK gluon tower may change our result by less than 10\%.

\begin{figure}[tb]
\begin{center}
\includegraphics[width=10cm]{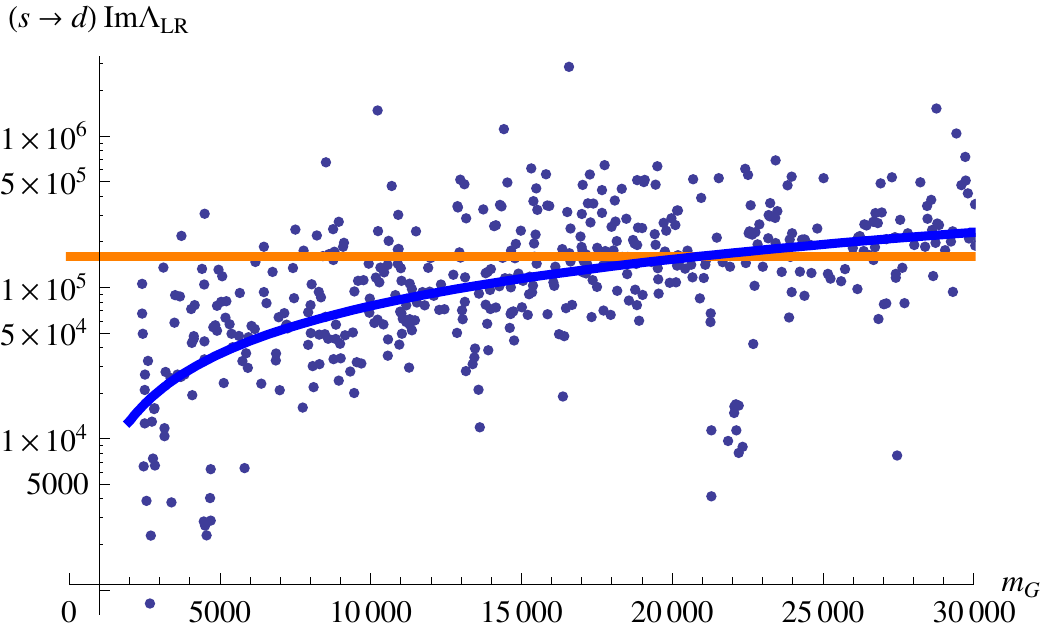}
\caption{Scan of the effective suppression scale of Im$C_{4K}$ in the RS model. All the points give the correct low-energy spectrum but most of the points with $m_G < 21$ TeV fail to satisfy the bound of $\Lambda>1.6 \cdot 10^5$ TeV. The blue line is a linear fit of the $M_G$ dependence. We have taken $c_{q_3} \in [0.4, 0.45]$, $c_{u_3} \in [-0.3, -.05]$, $c_{d_3}= -0.55$  and $|Y_*| \in [1,3] $. } \label{fig:RSbound}
\end{center}
\end{figure}

In order to understand  the actual flavor bound on the RS model in
more detail we have generated a sample of points with randomly
chosen values of $1/R'$ and brane Yukawa couplings of which we
selected 500 where the masses, the absolute values of the CKM
elements and the Jarlskog invariant approximately matches the SM
prediction. We then calculate the exact flavor suppression scale for
the $C_{4 K}$ operator. The result is presented in
Fig.~\ref{fig:RSbound}. We can say that, in accordance with our
analytical estimates, as long as $M_G$ is below 21 TeV the
majority of the generated points violate the flavor bound.
This turns the "coincidence problem" of RS \cite{APS} into a fine tuning problem: unless there
is some additional flavor structure one is likely to violate the
flavor bounds.

Note, that there is a model dependence that can strengthen or weaken the above obtained bound:
the matching of the bulk gauge coupling to the strong coupling can be changed by adding localized kinetic terms for the
gluon
\beq
\frac{1}{g_s^2(q)} = \frac{\log R'/R}{g_{s*}^2} + \frac{1}{g_{s,UV}^2(q)}
+ \frac{1}{g_{s,IR}^2(q)}
\eeq
A positive brane kinetic term would make the KK gluon more strongly coupled, which would make the flavor bounds more severe.
However, the UV brane coupling can be effectively negative at the TeV scale, if one includes the 1-loop running effects~\cite{gaugematching}~\footnote{We thank Kaustubh Agashe and Roberto Contino for pointing this out.},
\beq
\frac{1}{g_{s,UV}^2(q)}  = \frac{1}{g_{s,UV}^2(1/R)} -  \frac{b_3^{UV}}{8\pi^2} \log (1/q R)
\eeq
where $b_3^{UV}$ is the QCD beta functions of the zero modes localized around the UV branes. The running of the IR brane localized kinetic term is cut off
at the scale $\sim 1/R'$ therefore it will not involve a large logarithm and
we will neglect the IR brane localized terms, and focus only on the UV brane localized kinetic terms.
Assuming that the top is localized on the TeV brane, and all other fields on the UV brane for QCD we find $b_3^{UV}=8$.
Thus, the asymptotically free QCD running reduces the magnitude of $g_{s*}$ and thus the coefficient of the operators induced by the KK gluon exchange, while a bare UV brane localized kinetic term would enhance the effect.
Our 21 TeV bound presented above corresponds to a choice of boundary kinetic terms where the bare UV couplings exactly cancels the contribution from the running.
Another possibility  would be to assume no UV boundary kinetic term at the Planck scale.
In that case the coupling of the KK gluon is much weaker, changes from $g_{s*} \approx 6$ to $g_{s*} \approx 3$, and the bound is reduced by a factor of $\sim 2$ from 21 TeV to 10.5 TeV.
Yet another possibility is to pick a large bare UV coupling such that the bulk is strongly coupled, $g_{s*} \sim 4 \pi$.
This would enhance the flavor bound on the KK gluon mass by a factor of $\sim 2$ from 21 TeV to 42 TeV.
%Note, that here and in the following we have only included the contribution of one KK mode. Summing over the whole tower will further enhance the bound by about $\sim (\sum_1^\infty 1/n^2)^{\frac12}\sim 1.3$ to 27 TeV.

\vspace{1cm}

In the remainder of this paper we investigate how the flavor bounds
are modified in 5D models where the electroweak breaking sector
arises dynamically.

%%%%%%%%%%%%%%%%%%%%%%%%%%%%%%%%%%%%%%%%%%%%%%%%%%%%%%%%%%%%%%%%%%%
%%%%%%%%%%%%%%%%%%%%%%%%%%%%%%%%%%%%%%%%%%%%%%%%%%%%%%
\section{5D models for pseudo-goldstone Higgs}
\label{sec:pgh} \setcounter{equation}{0} \setcounter{footnote}{0}
%%%%%%%%%%%%%%%%%%%%%%%%%%%%%%%%%%%%%%%%%%%%%%%%%%%%%%
%%%%%%%%%%%%%%%%%%%%%%%%%%%%%%%%%%%%%%%%%%%%%%%%%%%%%%%%%%%%%%%%%%%

In this section we introduce the 5D framework of GHU and define particular models whose flavor structure  we will later study.
The basic ingredient of GHU  is the presence of a zero mode along the scalar ($A_z$) direction of the bulk gauge field.
This mode is identified with the SM Higgs boson.
The idea of GHU has a long history going back to Manton and Hosotani \cite{M}.
More recently, GHU has been formulated in 5D warped space-time \cite{OW,CNP,ACP} which, among other things, allows one to obtain a heavy enough top quark.
Other important developments are related to electroweak precision observables (EWPO's).
The S-parameter is within the experimental bounds if the KK scale of the theory is raised to 2-3 TeV \cite{ACP}.
The  $\rho$-parameter can be protected by custodial symmetry, and the discrete $L \leftrightarrow R$ symmetry protecting the $Zb\bar{b}$ vertex can be implemented \cite{CDP}.

The simplest GHU model with custodial symmetry is based on the $SO(5) \times U(1)_X$ (plus the color group) bulk gauge group
which is broken via boundary conditions to the SM group  $SU(2)_L \times U(1)_Y$ on the UV brane and to $SO(4) \times U(1)_X$
on the IR brane, where SO(4) $\sim$SU(2)$_L\times$SU(2)$_R$.
The fact that the SO(5)/SO(4) coset generators are broken on both branes results in four zero modes with the quantum numbers of the SM Higgs doublet.
At tree-level, these modes are massless due to 5D gauge invariance,  but at one loop they develop a potential.

We denote the dimensionful gauge coupling of SO(5) by $g_* R^{1/2}$.
A useful measure of the magnitude of  $g_*$ is the number of colors of the dual CFT by
\beq
\label{e.ncft}
N_{CFT} = \frac{16 \pi^2}{g_*^2}
\eeq
The 5D description is perturbative as long as $N_{CFT} \gg 1$.
We also allow for a UV brane localized gauge kinetic terms for $SU(2)_L$ (for simplicity, we set the $U(1)_Y$ brane kinetic term to zero) and parametrize it by $1/g_{UV}^2 = r^2 \log(R'/R)/g_*^2$.
The $SU(2)_L$ coupling is then related to the 5D parameters by
\beq
\label{e.wc} % weak coupling
g = {g_* \over \log^{1/2}(R'/R)\sqrt{1 + r^2}}
\eeq
For a small brane kinetic term, $r \ll 1$, and the Planck-TeV hierarchy, $\log (R'/R) \sim 37$, we need the bulk coupling $g_* \approx 4$ (corresponding to $N_{CFT} = 10$) in order to reproduce the SM weak coupling. With the brane kinetic term we have to make the bulk more strongly coupled, in particular for $r^2 = 4$ we go down to  $N_{CFT} = 2$.

The four components $h^a$ of the Higgs doublet are embedded into the gauge field $A_5$ as
\begin{equation}
A_z(z)=\sqrt{\frac{2}{R}} \frac{z}{R'} T_C^a h^a (x)
\end{equation}
Here, the $T_C^a$'s are the generators of the SO(5)/SO(4) coset. The
normalization is chosen such that $h^a$'s have the canonical kinetic terms in the 4D effective theory.
We fix the Higgs vev along the $h_4$ direction, and we denote $\la h_4 \ra = \ti v$.

The 5D model has two important scales that play a vital role in the dynamics.
One is $R'$ which sets the KK scale, and the masses of the lightest KK modes of electroweak gauge bosons are roughly $\sim 2.4 R'$.
Another important quantity is the global symmetry breaking scale $f_\pi$
(or the "Higgs decay constant") defined by
\begin{equation}
f_\pi=\frac{2}{g_* R'} = \frac{2}{g \sqrt{\log R'/R} \sqrt{1 + r^2} R'}.
\end{equation}
Because the logarithm is large, $f_\pi < R'$ by at least a factor of $\sim 2$.
The W-mass is connected to the scale $f_\pi$ and the Higgs vev via
\begin{equation}\label{equ:mw}
M_W^2 = \frac{g^2 f_\pi^2}{4} \sin^2 (\ti v/f_\pi)
\end{equation}
For $\ti v \ll f_\pi$, once recovers the SM formula $M_W^2 \approx g^2 \ti v^2/4$.

Fixing the gauge group still leaves several options for realizing the fermionic sector of the theory,
depending on how the SM quarks are embedded into SO(5) representations.
In this paper we consider three distinct realizations that have  previously appeared in the literature.
Before we move to the detailed description of the  model we point a few general model-building rules that we need follow.
\bi
\item {\it The SM quarks are identified with the zero modes of 5D quarks.}
The presence of zero modes depends on the boundary conditions.
There are two possibilities:
either we choose the right-handed chirality of the 5D fermion to vanish on a boundary (which is denoted as $[+]$),
or the left-handed chirality vanishes (denoted as $[-]$).
Left-handed zero modes - appropriate for SM doublet quarks - arise whent the right-handed chirality vanishes on both the UV and the IR brane, the choice denoted as $[++]$.
The $[--]$ boundary conditions lead to right-handed zero modes appropriate for the $SU(2)_L$ singlet SM quarks.
\item {\it Each of the SM quarks should be embedded in a separate SO(5) multiplet.}
In principle, SO(5) multiplets contain fields with the quantum numbers of both doublet and singlet quarks.
However, multiplets that yield both doublet left-handed zero modes and singlet right-handed
zero modes are problematic for the following reason. For the first
two generations, if the left-handed zero mode is localized close to  the
UV brane, then the right handed zero mode is localized at the IR brane,
which leads to problems with precision measurements.
For the third generation, the reason is more subtle and has to do with the
radiative generation of the Higgs potential; we will comment on that
later.
Thus, we need at least three SO(5) multiplets for each generation.
As a consequence, the field content of GHU models is necessarily larger than that of  RS.
\item
Apart from the zero mode, the remaining fields in the multiplets should yield only heavy KK modes.
One should take care that the boundary conditions do not lead to ultra-light KK modes.
This may happen for the quarks living in the same multiplet with UV localized zero modes, depending on the boundary conditions of the remaining fields.
The rule of thumb is that SO(5) partners of $[++]$ fields should be assigned $[-+]$ boundary conditions,
while the partners of $[--]$ fields should have $[+-]$.
\item
This zeroth-order picture is modified by the mass terms on the IR brane that mix different 5D multiplets.
These mass terms are necessary to arrive at  acceptable phenomenology.
First of all, since the Higgs field is a component of  $A_z$, it only couples 5D quarks from the same bulk multiplet.
In order to obtain non-zero quark masses, at least some of the zero modes should have non-vanishing components in more than one multiplet.
The boundary mass terms play a similar role as the Yukawa couplings in RS but, as we discuss later, there are some important differences.
\item At the end of the day,  the boundary conditions to SO(5) multiplets should  be such that {\it  SO(5) is broken on both the UV and the IR branes.}
The Wilson-line breaking is a non-local effect that is operating only when the gauge symmetry is broken on both endpoints of the fifth dimension.  If either the UV or the IR boundary conditions are SO(5) symmetric, the Wilson line can be rotated away, and the SM quarks do not acquire masses.
\ei

We move to discussing three specific realizations that satisfy the above requirements.

%%%%%%%%%%%%%%%%%%%%%%%%%%%%%%%%%%%%%%%%
\subsection{Spinorial}
\label{s.sp}
%%%%%%%%%%%%%%%%%%%%%%%%%%%%%%%%%%%%%%%%%

The spinor representation $\bf 4$  is the smallest SO(5) representation.
Although models with the third generation embedded in the spinorial representation have severe problems
with satisfying the precision constraints on the $Z b b$ vertex,
we do  include it in our study.
The reason is that this model is the simplest (it has the minimal number of bulk fields), and its flavor structure is most transparent.
Furthermore, almost identical flavor structure appears in the fully realistic models.

We consider 3 bulk SO(5) spinors for a single generation of quarks, $\Psi_q, \Psi_u,\Psi_d$ (recall that we omit the generation index; all fermionic fields should be read as three-vectors in the generation space).
Under the  SU(2)$_L\times$SU(2)$_R$ subgroup it splits as $4\to(2,1)+(1,2)$, so that an SO(5) spinor contains both
$SU(2)_L$ doublets and singlets.
Roughly, $\Psi_q$ will provide the zero mode for the left-handed quark doublets,
while $\Psi_u, \Psi_d$ for the right handed up and down-type quarks.
To obtain the SM zero mode spectrum we impose the following boundary conditions
\beq
\label{parities}
\Psi_q =  \bvec
q_q[+,+] \\ u_q^c[-,+] \\  d_q^c[-,+]
\evec
\qquad
\Psi_u =  \bvec
q_u[+,-]  \\ u_u^c[-,-] \\  d_u^c[+,-]
\evec
\qquad
\Psi_d =  \bvec
q_d[+,-] \\ u_d^c[+,-] \\  d_d^c[-,-]
\evec
\eeq
with the notation that the first component is a complete SU(2)$_L$ doublet,  while the lower two components are the two components of
an SU(2)$_R$ doublet, $q^c = (u^c,d^c)$.
Our model is similar to the one in ref. \cite{ACP}, even though we assign different IR boundary conditions for the $SU(2)_R$ doublets\footnote{In particular, $\Psi_q$ and $\Psi_u$ alone would be equivalent after interchanging $\ti M_u \to 1/\ti M_u$. With $\Psi_d$ included, the two models are not equivalent.}.

We denote the left-handed chirality modes of a Dirac field by $\chi$, while the right-handed chiralities by $\psi$.
For example $\chi_{q_q}$ stands for the left-handed chirality SU(2)$_L$ doublet contained in $\Psi_q$.
The above set of parity assignments  ensures the zero modes with SM quantum
numbers in $\chi_{q_q}, \psi_{u_u^c}$ and $\psi_{d_d^c}$.
However, at this point there would be no Yukawa couplings at all, since the
zero modes live in completely different bulk multiplets.
To obtain non-zero Yukawa couplings, at least some of the zero modes should have non-vanishing components in more
than one multiplet.
This can be achieved via the following IR localized mass terms:
\begin{equation}
\cl_{IR} = -
\left(\frac{R}{R'}\right)^4
\left [ \tilde{m}_u
\chi_{q_q} \psi_{q_u}
+ \tilde{m}_d \chi_{q_q} \psi_{q_d}
+ \tilde{M}_u ( \chi_{u_q^c} \psi_{u_u^c} +\chi_{d_q^c} \psi_{d_u^c} )
+ \tilde{M}_d  (\chi_{u_q^c} \psi_{u_d^c}+\chi_{d_q^c} \psi_{d_d^c} ) \right ]
\label{e.twistedBC}
\end{equation}
Here $\tilde{m}_{u,d},\tilde{M}_{u,d}$ are dimensionless 3 by 3 matrices,
which will play the similar role  as brane-localized Yukawa couplings in the original RS.
All flavor mixing effects in this model originate from the IR localized mass terms.
The effect of $\tilde{m}_u$ is to rotate the doublet zero mode partly into the $\Psi_u$ field, while
$\tilde{m}_d$ rotates it partly into $\Psi_d$.
At the same time $\tilde{M}_u$ will rotate the singlet up-type zero mode partly into $\Psi_q$, and similarly $\tilde{M}_d$ will rotate the down-type zero mode into $\Psi_d$.
The boundary mass terms respect the  SU(2)$_L\times$SU(2)$_R$ of the IR brane, but they break SO(5).
In the limit $\tilde{m}_u=\tilde{M}_u$ and $\tilde{m}_d=\tilde{M}_d$ SO(5) invariance is restored in the IR boundary conditions, and the zero  mode quarks become massless.

In the presence of the boundary terms the IR brane boundary conditions will be modified as
\bea
\label{e.ksbcir}
\psi_{q_q} &=& - \ti m_u \psi_{q_u} - \ti m_d \psi_{q_d}
\nn
\chi_{q_u} &=& \ti m_u^\dagger \chi_{q_q}
\nn
\chi_{q_d} &=& \ti m_d^\dagger \chi_{q_q}
\eea
\bea
\psi_{Q_q} &=& - \ti M_u \psi_{Q_u} - \ti M_d \psi_{Q_d}
\nn
\chi_{Q_u} &=& \ti M_u^\dagger \chi_{Q_q}
\nn
\chi_{Q_d} &=& \ti M_d^\dagger \chi_{Q_q}
\eea

%%%%%%%%%%%%%%%%%%%%%%%%%%%%%%%%%%%%%%%%
\subsection{Fundamental + Adjoint}
\label{s.ff}
%%%%%%%%%%%%%%%%%%%%%%%%%%%%%%%%%%%%%%%%%

As we explain later,  the model with fermions in the spinor representation turns out to have incurable problems:
the Higgs mass tends to be too light, and there is a large irreducible correction to the $Zb\bar{b}$ vertex.
The situation is improved in models where the doublet quarks are embedded in the fundamental representation of SO(5).
The original motivation for considering the fundamental represenation was the realization \cite{ACDP} that it is possible to greatly reduce the corrections to the $Z b\bar{b}$ vertex by using an embedding of the SM fermions into the custodially symmetric SU(2)$_L\times$SU(2)$_R$ model under which the $b_L$ is
symmetric under SU(2)$_L\leftrightarrow$SU(2)$_R$. The simplest
implementation of this $Z_2$ symmetry is when the left handed quarks
are in a bifundamental under SU(2)$_L\times$SU(2)$_R$, while $t_R$
is a singlet.

In the context of the SO(5) MCH model the minimal model is obtained
via introducing two fundamental ({\bf 5}) and one adjoint ({\bf 10})
representation.
This is the model for which the constraints from electroweak symmetry breaking and electroweak
precision constraints have been investigated in detail in~\cite{CPSW}.
Under U(1)$_X$ all bulk fermion carry charge -2/3.
Under the SO(4) subgroup $5\to (2,2)+(1,1)$, while $10\to
(3,1)+(1,3)+(2,2)$. In order to protect the $Zb\bar{b}$ vertex we
want $q_L$ to be part of (2,2), $t_R$ in (1,1) and $b_R$ in (1,3).
Thus the fermion content of this model will be   $2 \times 3$ 5D quarks in the fundamental
representation  and $1 \times 3$ quarks in the adjoint representation of $SO(5)$.
The fives (denoted $\Psi_{q}$, $\Psi_{u}$) each host three up quarks $u$, $\ti
u$, $u^c$, one down quark $d$ and one exotic charge $5/3$ quark $X$.
$q = (u,d)$ is hypercharge $1/6$ $SU(2)_L$ doublet, while $\ti q =
(X,\ti u)$ is hypercharge $-7/6$ $SU(2)_L$ doublet. $(q,\ti q)$ is a
bidoublet under $SU(2)_L\times SU(2)_R$. The tens (denoted $Q_{d}$)
hosts four up quarks $u^d$, $\ti u^d$, $u^l$, $u^r$, three down
quarks $d^d$, $d^l$, $d^r$,  and three exotics $X^d$, $X^l$, $X^r$.
These quarks are collected into an $SU(2)_L$ triplet $l =
(X^l,u^l,d^l)$,  an $SU(2)_R$ triplet $r = (X^r,u^r,d^r)$ and a
bidoublet $q^d,\ti q^d$, where $q^d = (u^d,d^d)$ is hypercharge
$1/6$ $SU(2)_L$ doublet, while $\ti q^d  = (X^d,\ti u^d )$ is
hypercharge $-7/6$ $SU(2)_L$ doublet.
Below we write the appropriate boundary conditions for the left-handed fields of every component:
\beq
\label{e.fac} %content
\Psi_{q} =  \bvec q_{q} [+,+] \qquad \ti q_{q}[-,+] \\
u_{q}^c[-,+] \evec \qquad
\Psi_{u} =  \bvec q_{u}[+,-] \qquad \ti
q_{u}[+,-] \\ u_u^c[-,-] \evec \eeq \beq \Psi_{d} =  \bvec l [+,-]
\qquad r = \bvec X_r[+,-] \\ u_r[+,-] \\ d_r[-,-] \evec
\\
q_d [+,-] \qquad \ti q_d [+,-] \evec
\eeq

With the above parity assignments we can also add IR boundary masses for the fermions, which are necessary to generate
the effective Yukawa couplings. These are given by
\beq
\label{e.fabm} % boundary masses
\cl_{IR} = -
\left(\frac{R}{R'}\right)^4 \left [ \ti m_u
(\chi_{q_{q}}\psi_{q_{u}} + \chi_{\ti q_{q}} \psi_{\ti q_{u}})  +
\ti M_u \chi_{u_{q}^c} \psi_{u_{u}^c} + \ti m_d ( \chi_{q_{q}}
\psi_{q_{d}} + \chi_{\ti q_{q}} \psi_{\ti q_{d}}) \right ] + \hc
\eeq
Note that the symmetries allow for one less boundary mass matrix than in the spinorial model.
The flavor structure of this model turns out to be identical to that of the spinorial model with $\ti M_d = 0$.

%%%%%%%%%%%%%%%%%%%%%%%%%%%%%%%%%%%%%%%%%%
\subsection{Four-Fundamental}
\label{s.fa}
%%%%%%%%%%%%%%%%%%%%%%%%%%%%%%%%%%%%%%%%%%%

Another possible implementation of the symmetry protecting the
$Zb\bar{b}$ vertex is to use four copies of bulk fundamentals for
every generation: two for the up-type quarks and two for the
down-type quarks \cite{CDP}. In the up sector one of the fundamental provides a
left handed doublet zero mode (in the bifundamental of the custodial
symmetry), and one right handed up-type singlet zero mode. In order to obtain the correct hypercharges for the SM quarks,
the U(1)$_X$ charge of these up-type fundamentals has to be -2/3.
To realize the down sector we need two additional fundamentals, with X charge 1/3, one providing
another doublet zero mode, and the other the down right zero mode.
In order to remove the additional doublet, we must assume that on
the Planck brane there is an additional right handed doublet, which
will marry one combination of the two doublet zero modes.

The boundary conditions are given by\footnote{Again, our IR boundary conditions are different than those of ref. \cite{CDP} but the physical content of the model is similar. In particular, the up-quark sector alone would be equivalent after replacing $\ti M_u$ with $1/\ti M_u$.}
\beq \Psi_{q_u} =  \bvec q_{q_u} [\pm,+] \qquad \ti q_{q_u}[-,+] \\
u_{q} [-,+] \evec \qquad \Psi_{u} =  \bvec q_{u}[+,-] \qquad \ti
q_{u}[+,-] \\ u_u[-,-] \evec \eeq \beq \Psi_{q_d} =  \bvec \ti
q_{q_d}[-,+]  \qquad q_{q_d} [\pm,+] \\ d_{q} [-,+] \evec \qquad
\Psi_d = \bvec \ti q_{d}[+,-] \qquad q_{d}[+,-] \\ d_d[-,-] \evec
\eeq
Here $[\pm]$ stands for mixed boundary conditions for the electroweak
doublets $q_{q_u}$ and $q_{q_d}$ on the UV brane:
\beq
\theta \chi_{q_{q_u}} - \chi_{q_{q_d}} = 0
\qquad
\psi_{q_{q_u}}   +  \theta  \psi_{q_{q_d}} = 0
\eeq where  $\theta$ is a $3 \times 3$  matrix that describes which combinations of the fields $\chi_{q_{q_u}}$ and
$\chi_{q_{q_d}}$ are removed on the UV brane.

Then the left handed zero modes from $[++]$ and the right handed
zero modes from $[--]$ fields are all elementary.  We again add the
IR boundary mass terms
\bea -\left(\frac{R}{R'}\right)^4   \left [
\ti m_u (\chi_{q_{q_u}} \psi_{q_{u}} + \chi_{\ti q_{q_u}} \psi_{\ti
q_{u}}) + \ti M_u \chi_{u_{q}}\psi_{u_{u}} +\ti m_d (\chi_{q_{q_d}}
\psi_{q_{d}} + \chi_{\ti q_{q_d}} \psi_{\ti q_{d}}) + \ti M_d
\chi_{d_{q}}\psi_{d_{d}} \right ] + \hc \nonumber \\
\eea
We have four IR matrices, just like in the spinorial model.
In this model, however, there is  an additional source of flavor violation - the
matrix $\theta$ that sets  the UV boundary conditions.

%%%%%%%%%%%%%%%%%%%%%%%%%%%%%%%%%%%%%%%%%%%%%%%%%%%%%%%%%%%%%%%%%%%%%%%%%%
%%%%%%%%%%%%%%%%%%%%%%%%%%%%%%%%%%%%%%%%%%%%%%%%%%%%%%%%%%%%%%%%%%%%%%%%
\section{Higgs potential}
\label{sec:hp} \setcounter{equation}{0} \setcounter{footnote}{0}
%%%%%%%%%%%%%%%%%%%%%%%%%%%%%%%%%%%%%%%%%%%%%%%%%%%%%%
%%%%%%%%%%%%%%%%%%%%%%%%%%%%%%%%%%%%%%%%%%%%%%%%%%%%%%

In this section we review the computation of the one-loop Higgs potential in the GHU models considered in this paper.
Determining the shape of the potential will allow us to pinpoint the regions of parameter space that lead to a correct electroweak breaking vacuum.
We will later use this input in our studies of the parameter space allowed by flavor constraints.
This section is more technical and slightly outside the main line of the paper, yet we include it to keep the paper self-contained.
Those readers whose  primary interest is in flavor physics are cordially invited to jump straight to the next section.

A radiative Higgs potential is generated at one-loop level  because
the tree-level KK mode masses depend on the vev $\ti v$ of the
Wilson line. The simplest way to calculate is to use the so-called
spectral function $\rho(p^2) = \det (-p^2 + m_n^2(\ti v))$, that is
a function of 4D momenta whose zeros encode the whole KK spectrum in
the presence of the electroweak  breaking. With  a spectral function
at hand, we can compute the Higgs potential from the
Coleman-Weinberg formula, \beq
\label{e.cws0} % Coleman Weinberg spectral
V(\ti v) = {N \over (4 \pi)^{2}}  \int_0^\infty dp p^{3} \log \left ( \rho[-p^2] \right )
\eeq
where $N = - 4 N_c$ for quark fields and $N = + 3$ for gauge bosons.
The spectral function can be computed by solving the equations of motion and the boundary conditions in the presence of the Higgs vev.

The leading contribution to the Higgs potential comes from the top
quark sector and the gauge sector. Thus, we can restrict to
computing the spectral functions of the top quark KK tower
($\rho_t$), the W boson tower ($\rho_W$) and the  Z boson tower
($\rho_Z$).
For the sake of this computation we ignore the mixing of the top quark with the first two generations.
An explicit expression for the potential in terms of
these spectral functions is (with $t=p^2$):
\beq
\label{e.cws} % Coleman Weinberg spectral
V(\ti v) = \frac{3}{32 \pi^2}  \int_0^\infty dt  t \left[
-4 \log\rho_t(-t) +2 \log \rho_W (-t) +\log \rho_Z (-t) \right].
\eeq

The gauge sector is common to all three models.
The spectral functions for the SO(5) GHU model were already given in detail in~\cite{FPR,MSW}.
For completeness we will summarize the results below.
Of the various SO(5) gauge bosons only the masses of the tower corresponding to the W,Z bosons depends on the Higgs VEV.
The spectral function has the form
\beq
\rho_{W,Z}(-p^2)=1 + f_{W,Z}(-p^2) \sin^2(\ti v/f_\pi),
\eeq
where the form factors $f^{W,Z}$ do not depend on $\ti v$.
The equations of motion in AdS are solved in term of the Bessel functions, and the form factors turn out to be complicated combinations thereof.
There is however a way to organize them in a more convenient form by using the generalized warped-space trigonometric functions introduced in ref.~\cite{AA}.
We define $C(z)$ and $S(z)$ to be the two independent solution of the equations of motion that satisfy the UV boundary conditions
$C(R) = 1$, $C'(R) = 0$, $S(R) = 0$, $S'(R) = m$.
Explicitly they read,
\begin{eqnarray}
\label{e.csads}
C(z) &=& {\pi m z \over 2}  \left [
Y_0 \left ( m  R \right )      J_1 \left ( m z \right )
- J_0 \left ( m R \right )     Y_1 \left ( m z \right )
\right ]
%\nn
%C'(R') &=& \frac{\pi m^2 R'^2 }{ 2 R}  \left [ Y_0 \left ( m  R \right )      J_0 \left ( m R' \right ) - J_0 \left ( m R \right )     Y_0 \left ( m R' \right ) \right ]
\nn
S(z) &=&  {\pi m  z \over 2 }  \left [
- Y_1 \left ( m R \right )     J_1 \left ( m z \right )
+ J_1 \left ( m R \right )      Y_1 \left ( m z \right )
\right ]
%\nn S'(R') &=& \frac{\pi m^2 R'^2 }{ 2 R}   \left [ - Y_1 \left ( m R \right )     J_0 \left ( m R' \right ) + J_1 \left ( m R \right )      Y_0 \left ( m R'\right ) \right ]
\end{eqnarray}
Using this, the form factors can be abbreviated to
\bea
f_{W}(m^2) &=&
\frac{m}{2} \left( \frac{R'}{R}\right) \frac{1}{[C'(R') - r^2 m R \log(R'/R)   S'(R')]  S(R')}
\nn
f_Z(m^2)  &=& \frac{m}{2} \left( \frac{R'}{R}\right) \frac{1 + {\tan^2 \theta_W \over 1 + r^2} \left( 1 -r^2  m R\log(R'/R) {S'(R') \over C'(R')}\right ) }{[C'(R')- r^2  m R \log(R'/R) S'(R')]  S(R')}
\eea
where $\tan \theta_W = g'/g$.
In the evaluation of the Higgs potential we use the warped trigonometric functions at $m^2 = - p^2$, e.g.
\bea
%C'(R')&=&
% p^2 R' \left [
% K_{0} \left (p R\right )      I_{0} \left ( p R' \right )
% -I_{0} \left ( p R\right )     K_{0} \left ( p R' \right ) \right ]
% \nn
S(R') &=& i p R' \left [
   - I_{1} \left ( p R \right )      K_{1} \left ( p R' \right )
       +  K_{1} \left ( p R \right )     I_{1}  \left ( p R' \right )
       \right ].
       \eea

We move to the top sector.
Again, the equations of motions can be solved in terms of the Bessel functions,
but there is also a dependence on the bulk mass parameters $c$.
As for the gauge bosons, it is convenient to introduce the AdS warped trigonometric functions
(here already evaluated for $m^2 = - p^2$)
\bea C_c &=& {p R }
\left(\frac{R}{R'}\right)^{-c-1/2}\left [
K_{c-1/2} \left (p R\right )      I_{c+1/2} \left ( p R' \right )
+ I_{c-1/2} \left ( p R\right )     K_{c+1/2} \left ( p R' \right )
\right ] \nn S_c &=&  p R \left(\frac{R}{R'}\right)^{-c-1/2}\left [
- I_{c+1/2} \left ( p R \right )      K_{c+1/2} \left ( p R' \right )
+  K_{c+1/2} \left ( p R \right )     I_{c+1/2}  \left ( p R' \right
) \right ]
\eea

The form of the spectral function depends on the fermion representations, and we have to treat each of the three models separately.
In the {\bf spinorial} model,
the spectral function can then be parameterized in terms of the form factors $f_{2,4}$
\beq
\label{e.kss} %spectral
\rho_t(-p^2) =   1 + f_2^t(-p^2) \sin^2(\ti v/2f_\pi)  + f_4^t(-p^2)
\sin^4(\ti v/2f_\pi) \eeq Note that the spectral depends on
$\sin^2(\ti v/2f_\pi)$ rather than $\sin^2(\ti v/f_\pi)$, which is a
peculiarity of the spinorial representation. The form factors can be
written as \beq f_2^t  = {F_2^t \over F_0^t} \qquad f_4^t  = {F_4^t
\over F_0^t} \eeq \bea F_0(p^2) = & \left [S_{-c_q} C_{c_u} + |\hat
m_u^2| S_{-c_u} C_{c_q} + |\hat m_d^2| S_{-c_d} {C_{c_u}C_{c_q}
\over C_{c_d}} \right ] \times \nn  & \left [S_{c_u} C_{-c_q} +
|\hat M_u^2| S_{c_q} C_{-c_u} - |\hat M_d^2| S_{-c_d} {S_{c_u}
S_{c_q} \over C_{c_d}} \right ] \nn F_2(p^2) = & - |\hat m_u - \hat
M_u|^2 + (|\hat m_u|^2 - |\hat M_u|^2)(S_{c_q} S_{-c_q} - S_{c_u}
S_{-c_u}) \nn & - (|\hat m_d|^2 - |\hat M_d|^2) S_{c_u} S_{-c_d}
{C_{c_u} \over C_{c_d}} + (|\hat m_d|^2 |\hat M_u|^2 - |\hat m_u|^2
|\hat M_d|^2 ) S_{c_q} S_{-c_d} {C_{c_q} \over C_{c_d}} \nn F_4(p^2)
= &|\hat m_u - \hat M_u|^2 \eea The hatted boundary masses are again
defined as \beq \hat m_u = (R'/R)^{c_u-c_q} \ti m_u \quad \hat M_u =
(R'/R)^{c_u-c_q} \ti M_u \quad \hat m_d = (R'/R)^{c_d-c_q} \ti m_d
\quad \hat M_d = (R'/R)^{c_d-c_q} \ti M_d \eeq

In the model with {\bf fundamentals + adjoint} the spectral functions read
\beq
\label{e.fas} %spectral
\rho_t(m^2) =  1 + f_2^t (m^2) \sin^2(\ti v/f)  +  f_4^t  \sin^4(\ti
v/f). \eeq
Note that for the fundamental and adjoint representations the spectrals are expressed in terms of  $\sin^2(\ti v/f)$,
rather than $\sin^2(v/2f)$ as for the spinors.
We write
\beq f_2^t = {F_2^t(m^2) \over F_0^t(m^2)} \quad f_4^t =
{F_4^t(m^2) \over  F_0^t(m^2)} \eeq \bea & F_0^t = \left [S_{-q}
C_{u} +  |\hat m_u|^2 C_{q} S_{-u} + {S_{-d} \over C_{d}} |\hat
m_d|^2 C_{q} C_{u} \right ] \left [C_{-q} S_{u} + |\hat M_u|^2 S_{q}
C_{-u} \right ] \nl F_2^t = {1 \over 2} \left ( - |\hat m_u - \hat
M_u|^2 + (|\hat m_u|^2 - |\hat M_u|^2)(2 S_{q} S_{-q} - S_{u}
S_{-u}) \bnl - {S_{-d} \over C_{d}} |\hat m_d|^2 (S_{u} C_{u} + 2
|\hat M_u|^2 S_{q} C_{q} ) \bnl + { |\hat m_u|^2 (|\hat m_u|^2 -
|\hat M_u|^2) S_{q} S_{-u} + {S_{-d} \over C_{d}}  |\hat m_u|^2
|\hat m_d|^2 S_{q} C_{u} \over C_{-q} C_{u} -  |\hat m_u|^2 S_{q}
S_{-u} - {S_{-d} \over C_{d}} |\hat m_d|^2 S_{q} C_{u} } \right )
\nl F_4^t = {1 \over 2} |\hat m_u - \hat M_u|^2 \eea

Finally, in the model with {\bf four fundamental} multiplets
\beq
\label{e.mfs} %spectral
\rho_t(m^2) =  1 + f_2^t (m^2) \sin^2(\ti v/f)  +  f_4^t  \sin^4(\ti
v/f) \eeq
\beq f_2^t = {F_2^t(m^2) \over  F_0^t(m^2)} \quad f_4^t =
{F_4^t(m^2) \over  F_0^t(m^2)} \quad F_0^t(m^2) = F_0^0 + |\theta|^2
F_0^\theta \quad F_2^t(m^2) = F_2^0 + |\theta|^2 F_2^\theta \eeq
We find
%\left [C_{-q_d} C_{d} - |\hat m_d^2| S_{q_d} S_{-d} \right ]
%\left [C_{-q_u} C_{u} - |\hat m_u^2| S_{q_u} S_{-u} \right ]
\bea F_0^0 = &
\left [S_{-q_u} C_{u} + |\hat m_u^2| C_{q_u} S_{-u} \right ]
\left [C_{-q_u} S_{u} + |\hat M_u^2| S_{q_u} C_{-u} \right ]
%\nn \cdot &
\nn F_0^\theta = & \left [C_{-q_u} S_{u} + |\hat M_u^2| S_{q_u}
C_{-u} \right ] \left [C_{-q_u} C_{u} - |\hat m_u^2| S_{q_u} S_{-u}
\right ]  \nn & \times \left [S_{-q_d} C_{d} + |\hat m_d^2| C_{q_d}
S_{-d} \right ] / \left [C_{-q_d} C_{d} - |\hat m_d^2| S_{q_d}
S_{-d} \right ]
\nn F_2^0 = &  % \nn \cdot &
{1 \over 2} \left ( - |\hat m_u - \hat M_u|^2 + (|\hat m_u|^2 -
|\hat M_u|^2) (2S_{q_u} S_{-q_u} - S_{u} S_{-u}) \right ) \nn + & {1
\over 2}  |\hat m_u|^2 (|\hat m_u|^2 - |\hat M_u|^2) S_{q_u} S_{-u}/
\left [C_{-q_u} C_{u} - |\hat m_u^2| S_{q_u} S_{-u} \right ] \nn
F_2^\theta = & (|\hat m_u|^2 - |\hat M_u|^2)C_{-q_u} S_{q_u}  \nn &
\times \left [S_{-q_d} C_{d} + |\hat m_d^2| C_{q_d} S_{-d} \right ]
/ \left [C_{-q_d} C_{d} - |\hat m_d^2| S_{q_d} S_{-d} \right ] \nn
F_4^t = & {1 \over 2} |\hat m_u - \hat M_u|^2 \eea

These expressions can be used for the calculation of the Higgs
potential in each model. One practical option is to numerically
calculate the integral in (\ref{e.cws}). However, it is useful to
gain some more insight into the shape of  the potential. This can be
done by expanding the potential in powers of $x \equiv \sin^2(\ti
v/f_\pi)$ (for the top contribution in the spinorial representation
we expand in $x = \sin^2(\ti v/2f_\pi)$). One complication in this
expansion is that the integral in (\ref{e.cws}) is logarithmically
IR divergent in the limit $x \to 0$. This is due to the fact that
some of the masses vanish in the limit $\ti v\to 0$, and the usual
CW formula contains terms of the form $m^4 \log m^2$.

\begin{figure}[tb]
\begin{center}
\includegraphics[width=8cm]{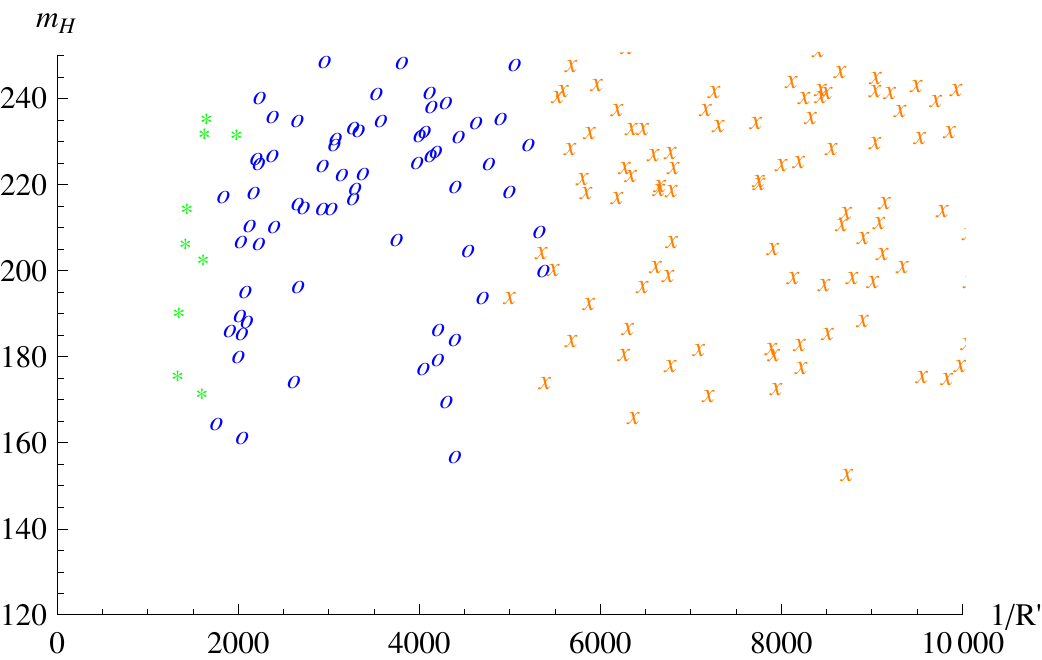}\quad \includegraphics[width=7.5cm]{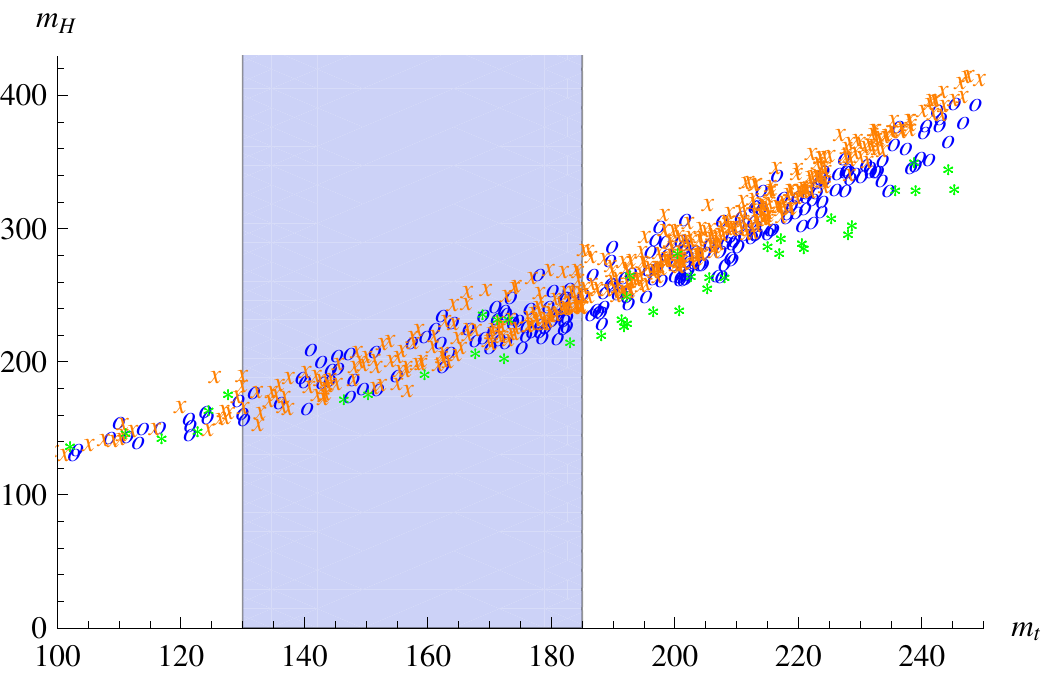}
\end{center}
\caption{Left panel: the dependence of the Higgs mass on $1/R'$ for
a variety of input parameters that give successful electroweak
symmetry breaking for the model with bulk fundamentals and adjoints
with points selected to give the top mass in the physical range.
Right panel: the correlation between the top and the Higgs masses
for the same case.  Orange (x), blue (o), and green (*) points
correspond to $\epsilon <0.1,0.1<\epsilon< 0.3$ , and
$0.3<\epsilon$. } \label{fig:mhplotsFA}
\end{figure}

\begin{figure}[tb]
\begin{center}
\includegraphics[width=8cm]{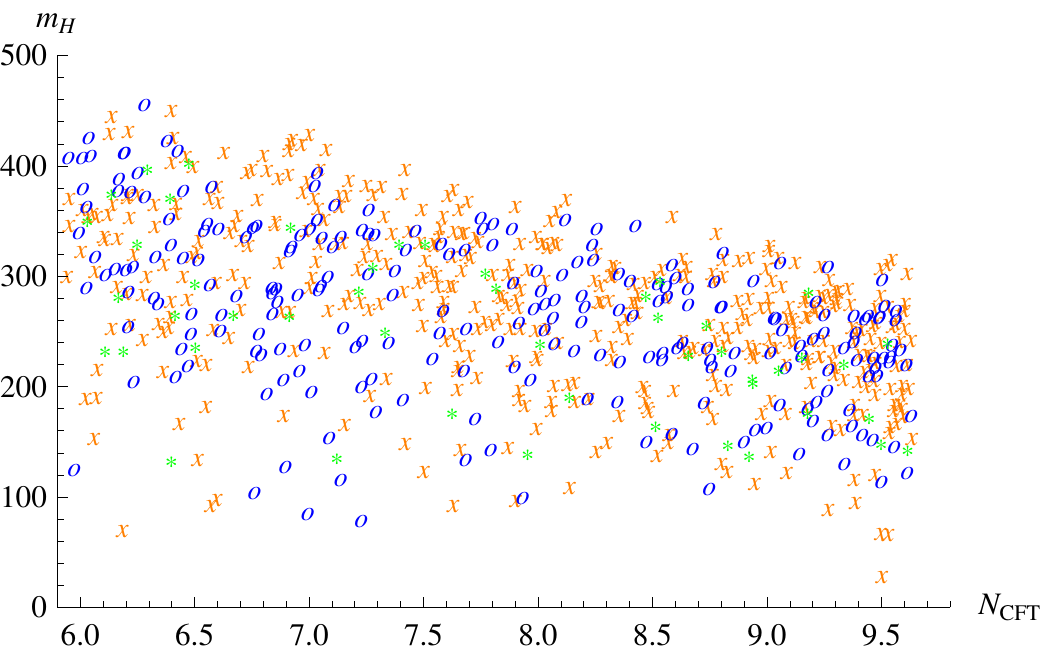}
\caption{The dependence of the Higgs mass on the effective number of
colors $N_{CFT}$ (corresponding strength of the bulk coupling) for
the case of bulk fundamental plus adjoint fermions. Orange (x), blue (o), and green (*) points
correspond to $\epsilon <0.1,0.1<\epsilon< 0.3$ , and
$0.3<\epsilon$.}
\label{fig:NvsRp}
\end{center}
\end{figure}

The top contribution to the Higgs potential has the form
\begin{equation}
V_t(x)= -\frac{3}{8 \pi^2} \int_0^\infty dt \, t \log [ 1 + f_2^t(-t) x +f_4^t(-t) x^2 ]
\end{equation}
For small $x$ this can be expanded as
\begin{equation}
V_t(x)= a_1^t x +a_2^t x^2 +n_2^t x^2 \log\frac{2 c_t x}{\Lambda^2} + \co(x^3)
\end{equation}
We defined the coefficient $c_t$ that captures the IR behavior of
the form factors,  namely $f_1^t(-t) \approx -f_2^t(-t) \approx
c_t/t$ for $t \to 0$, and it is related to the SM top quark mass by
$m_t^2 \approx c_t x$. This coefficient can be read off from the top
spectral function by replacing $C_c \to 1$, $S_c \to i t R'
(R'/R)^{-2c} f_c^{-2}$. We can then show that the general expression
for the coefficients in the expansion should be
\begin{eqnarray}
&& a_1^t= -\frac{3}{8 \pi^2} \int_0^\infty dt \, t f_2^t \nonumber \\
&& a_2^t=  -\frac{3}{16 \pi^2} \left [
\int_0^\infty dt \, t \left(2 f_4^t - (f_2^t)^2 +\frac{c^2}{\Lambda^4 \sinh^2(t/\Lambda^2)}\right)
-\frac{3}{2} c_t^2 \right ]
\nonumber \\
&& n_2^t= -\frac{3}{16 \pi^2} c_t^2
\end{eqnarray}
The scale $\Lambda$ is an IR regulator and it may take an arbitrary
value. The expansion parameters depend on $\Lambda$ in such a way
that the dependence on the  IR regulator cancels out at order $x^2$.

In the cases when the top contribution, dominates the minimum of the
potential is given by the approximate formula: \beq x_m = - {a_1^t
\over 2 a_2^t + n_2^t + 2 n_2^t \log (2 c x_m/\Lambda^2)} \eeq In
practice, however, the acceptable minimum with $x_m \ll 1$ occurs
only for fine-tuned  values of the parameters such that $a_1^t$ is
much smaller than its natural value $\sim m_t^2 (R')^2/ 4 \pi^2$. In
that case, the W and Z boson contributions can significantly shift
the minimum, and the acceptable EW breaking vacuum occurs for
different c-parameters than without the gauge contributions. The
gauge contributions are calculated from \bea V_W(x) &=& \frac{3}{16
\pi^2} \int_0^\infty dt \, t \log [ 1 + f_2^W(-t) x  ] \nn V_Z(x)
&=& \frac{3}{32 \pi^2} \int_0^\infty dt \, t \log [ 1 + f_2^Z(-t) x
] \eea and the expansion can be done analogously as for the top, by
setting $f_4^{W,Z} = 0$. In our numerical studies we take the gauge
contributions into account.

We turn to discussing the main features of the Higgs potential
generated by the top, bottom and gauge loops.  We have scanned the
parameter space to find self-consistent combinations leading to
realistic EWSB. In the scan we first choose the brane kinetic term
$r$, the bulk masses $c_{q_3}$ and $c_{d_3}$, the size of the brane
masses $\ti m_u$, $\ti m_d$, $\ti M_u$, $\ti M_d$, the KK scale
$1/R'$ and keep the hierarchy fixed at $R/R'= 10^{-16}$. We then
determine $v/f_\pi$ such that the $m_W$ mass is reproduced and
finally choose $c_{u_3}$ such that the minimum of the
Coleman-Weinberg potential is really at this value of $v/f_\pi$.

For the case of fundamental plus adjoint bulk fermions we do find
plenty of realistic values for the Higgs and the top masses, in
agreement with the results of~\cite{MSW}. This is illustrated in
Fig.~\ref{fig:mhplotsFA}. Note however, that in order to fit the
W-mass successfully with a sufficiently small $\epsilon$ one is
usually forced to introduce UV localized kinetic terms. This is
illustrated in figure~\ref{fig:NvsRp}. For our analysis of the
flavor scales we are only selecting from the points that give a
satisfactory top with low values of $\epsilon$.

\begin{figure}[tb]
\begin{center}
\includegraphics[width=7.cm]{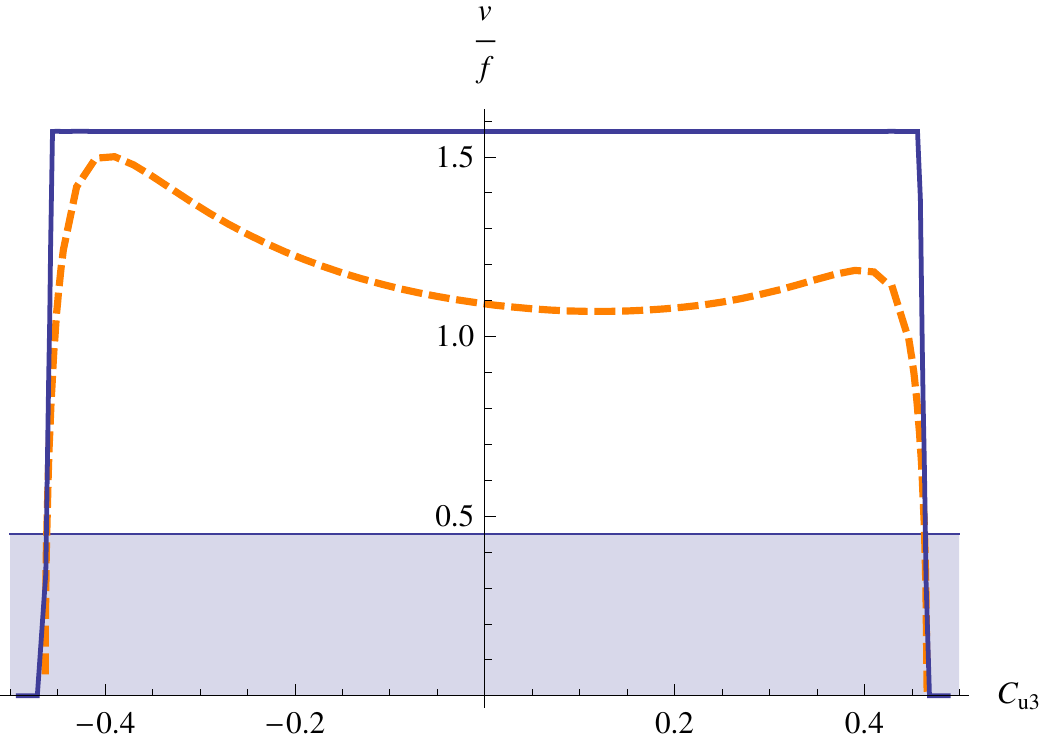} \quad
\includegraphics[width=7.cm]{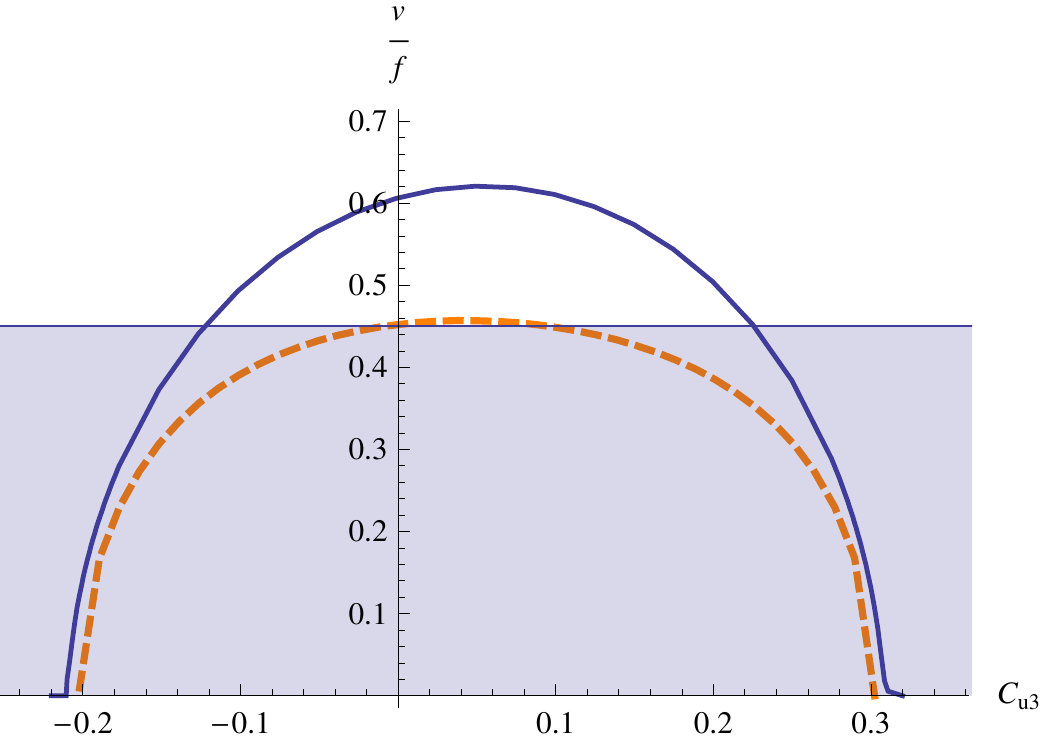}
\caption{The plots show the dependence of $\ti v/f_\pi$ on  $c_u$ of the
third generation for two sets of parameters in the model with
fundamentals and adjoints. The first example on the left shows a
generic point where there are only two narrow regions of $c_u$ that
lead to an acceptable electroweak breaking vacuum. The continuous
line shows the result of the minimization of the full potential and
the dashed line shows our approximation for regions where $\sin \ti
v/f_\pi$ is small.  In the left plot for $-0.46 \ltap c_{u_3} \ltap 0.46 $  the minimum of the potential is at $\ti v = \pi f_\pi/2$,
while for $ c_{u_3} \ltap -0.47 $ and $ c_{u_3} \gtap 0.47 $ the
minimum is at $\ti v = 0$. In the case of $\ti v = \pi f_\pi/2$,
electroweak symmetry is maximally broken $m_W = g f/2$, whereas for
$\ti v = 0$, electroweak symmetry is unbroken ($m_W = 0$).
In both cases there are massless fermions in the spectrum. For
$-0.47 \ltap c_{u_3} \ltap -0.46 $ and  $0.46 \ltap c_{u_3} \ltap 0.47 $ there is a minimum at intermediate values of $\ti v/f_\pi$, yet
an  additional tuning is required to arrive at $\ti v/f_\pi \ll 1$. For
$\ti v/f_\pi < 0.45$ we find $-0.464 < c_{u_3} < -0.462 $ and  $0.464 < c_{u_3} < 0.466 $. In this case one needs to tune $c_u$ to more than
a percent level to get successful electroweak symmetry breaking. The
right plot shows the same for carefully chosen values of the input
parameters, where the local tuning is more modest. In this best case
scenario we can have a realistic electroweak symmetry breaking
minimum for the region $-0.21 < c_{u_3} < -0.13 $ or $0.22 <
c_{u_3} < 0.31 $. The parameters chosen for this plot are
$R'/R=10^{16}$,$1/R'=1.5$ TeV and $c_{Q_3}= 0.42$, $c_{d_3}= -0.56$,
$\ti m_u =5$, $\ti m_d =1$, $\ti M_u =0$, $r=1$ (right: $c_{Q_3}= 0.1$,
$c_{d_3}= -0.56$, $\ti m_u =1$, $\ti M_u =-0.5$, $r=0.47$). }
\label{fig:tunecu}
\end{center}
\end{figure}

However, there is a fine tuning reminiscent of the little hierarchy
problem showing up: if one fixes the mixing parameters
$\tilde{m}_{u,d}, \tilde{M}_{u,d}$ and the bulk masses $c_{q,d}$,
then for given radii $R,R'$ and for generic choices of the other
parameters we have only a very narrow region in the parameter $c_u$
that produces an $\epsilon$ which is phenomenologically acceptable
(for example $0.1<\epsilon<0.45$). This suggests that in the
interesting region with proper electroweak symmetry breaking there
will be a strong sensitivity of $\epsilon$ to the input parameters.
This was first pointed out in~\cite{Panico:2008bx} and we find it to
be  a general property of all representations studied in this paper.
To illustrate this we show two examples of the dependence of
$\epsilon$ on $c_u$ in Fig.~\ref{fig:tunecu}. The first one is a
randomly chosen point where one has to adjust $c_u$ to a high
precision in order to find proper EWSB, while the second one
corresponds to the best case scenario that we could find after
searching for regions where the tuning is milder. We can see that
the derivative of $\epsilon$ is very large in the first point, and
even in the second point it is still quite sizeable.

\begin{figure}[tb]
\begin{center}
\includegraphics[width=8cm]{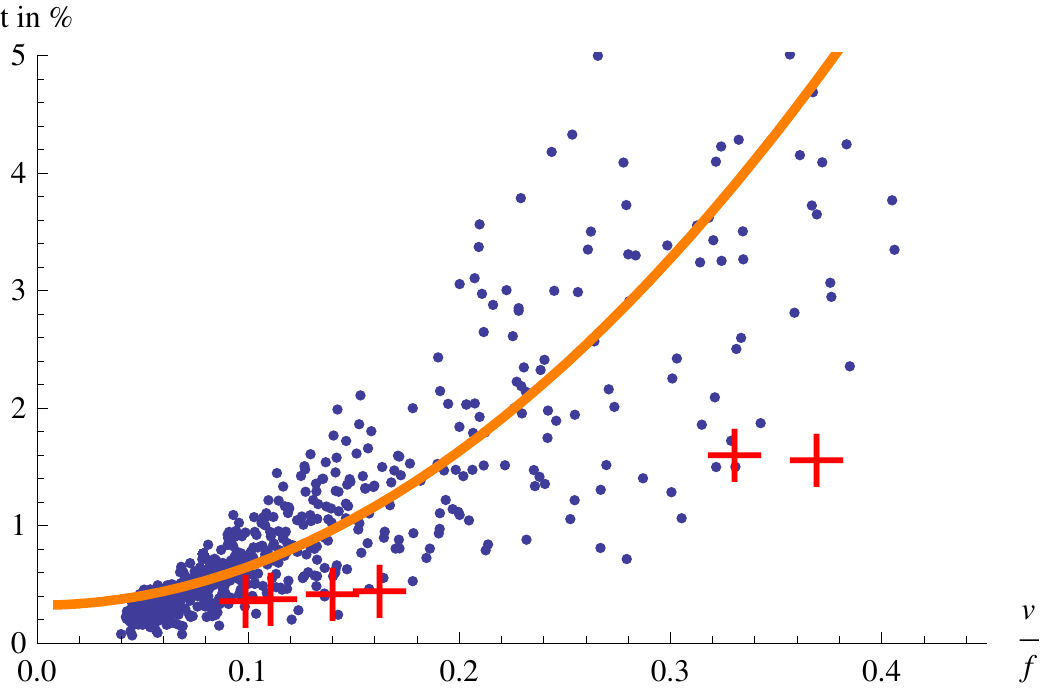}
\end{center}
\caption{This plot shows the local tuning $t$ for the model with
adjoints and fundamentals. The orange line is a fit to a quadratic
polynomial for which we find $t \sim \frac14
\epsilon^2$.} \label{fig:tuning}
\end{figure}

In order to quantify the tuning in these models we have calculated
the local sensitivity to the input parameters in the ranges given in
Sec.~\ref{sec:numerics}, for the regions with acceptable electrowaek
symmetry breaking only. We then define the local fine tuning
as~\cite{BG,AC} \beq t^{-1} = {\rm max} \left| \frac{\partial \log
\epsilon}{\partial \log a_i}\right| \eeq In this scan we have only
included points where EWSB happens with a sufficiently small S
parameter and excluded all other points. The parameters $a_i$ were
taken to be $c_{q_3}, c_{u_3}, \ti m_u, \ti M_u$. We then estimate
the average tuning for given $\epsilon$ by fitting the average of
$t$ with a quadratic polynomial in $\epsilon$. The best fit is
approximately \beq t \sim \frac14 \epsilon^2 \eeq which
qualitatively agrees with the $\epsilon^2$ estimate of~\cite{ACP}
but is numerically somewhat stronger. This implies that the average
local tuning is a about half a percent for $\epsilon =0.1$, while
about $5\%$ for $\epsilon =0.4$. To verify those estimates we
extended the parameter space to a large grid in the parameters
$c_{q_3}, c_{u_3}, \ti m_u, \ti M_u$ and checked that the above fit
for $t=t(\epsilon)$ remains a conservative lower bound over all
parameter space. These averages are shown as crosses in
Fig.~\ref{fig:tuning} where each cross contains the average of about
200 points. Note however that one can find restricted local areas in
parameter space where the fine tuning is less severe than the above
quoted average.

\begin{figure}[tb]
\begin{center}
\includegraphics[width=8cm]{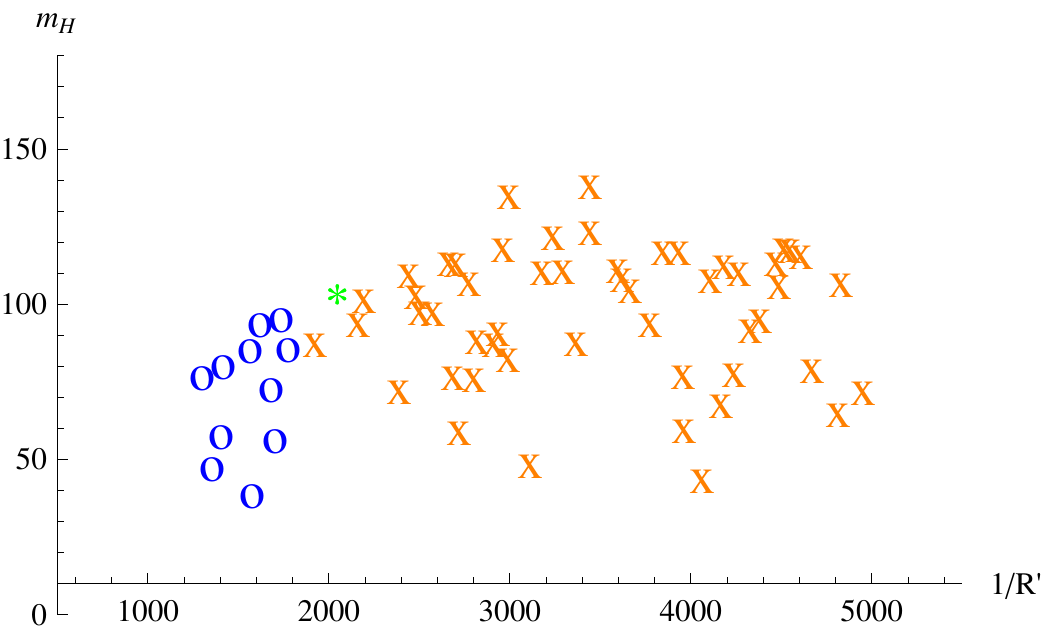}\quad \includegraphics[width=7.5cm]{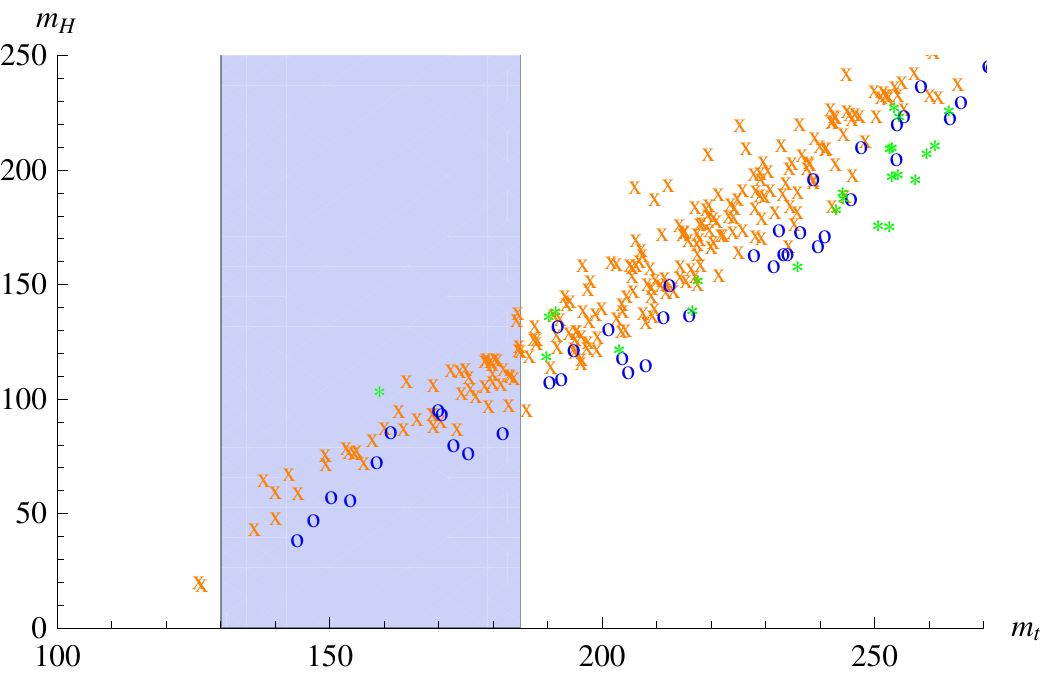}
\end{center}
\caption{ Left panel: the dependence of the Higgs mass on $1/R'$ for
a variety of input parameters in the model with spinors that give successful electroweak
symmetry breaking with points selected to give the top mass in the physical range. Right panel: the
correlation between the top and the Higgs masses. Orange (x), blue (o), and green (*) points correspond to $\epsilon <0.3,0.3<\epsilon< 0.4$ , and $0.4<\epsilon$. }
\label{fig:mhplots}
\end{figure}

For the  {\it spinorial representations} we find that  the bottom KK
tower plays an important role in electroweak symmetry breaking.
Without including the bottom tower we could not find any point with
a sufficiently heavy Higgs mass. Including the bottom improves the
situation because for the spinorial representation there is also a
light KK mode in the bottom sector.\footnote{We thank Roberto
Contino for discussions on this point.} However, it is exactly this
state that is responsible for the large shift in the $Zb\bar b$
vertex. The fine tuning in the top sector is somewhat stronger for
the spinorial case than for the fundamental+adjoint discussed above.
Once we impose the fine-tuning in the input parameters of the
theory, we still need to make sure that the Higgs is sufficiently
heavy. Most of the time the top is too heavy. The reason behind this
is that there is a very strong correlation between the Higgs and the
top masses in this model. These correlations between the Higgs mass
and $1/R'$ and $m_{top}$ are summarized in Fig.~\ref{fig:mhplots}.

%%%%%%%%%%%%%%%%%%%%%%%%%%%%%%%%%%%%%%%%%%%%%%%%%%%%%%%%%%%%%%
\subsection{Constraints from Electroweak Precision Tests}
%%%%%%%%%%%%%%%%%%%%%%%%%%%%%%%%%%%%%%%%%%%%%%%%%%%%%%%%%%%%%%

Apart from  yielding the correct electroweak breaking vacuum,
phenomenologically acceptable GHU models must pass stringent electroweak precision tests \cite{AC,CPSW}.
Since SO(5) models are endowed with custodial symmetry, there is no tree-level constraints from the T parameter.
The S parameter however is an issue.
The expression for the S parameter (for vanishing brane kinetic terms) is given by \cite{ACP}
\beq
S = 4 \pi v^2 \left [
{(\int_{R}^{R'} a )(\int_{R}^{R'} a^{-1})^2
- \int_{R}^{R'} a  (\int_{z}^{R'} a^{-1} )^2 \over \int_{R}^{R'} a^{-1} }
\right]
\approx  {3 \pi v^2 R'{}^2 \over 2}
\eeq
Demanding $S < .2$ yields the bound $R'^{-1} > 1.2 \tev$.
This implies that the lightest gauge boson KK modes have masses $\sim 2.4 R'^{-1} \sim 3 \tev$.

Another potentially dangerous contributions to the electroweak observables are the corrections to the
$Z b_L \ov b_L$ coupling that is measured with the $.25 \%$ accuracy.
There are two potential sources of deviations in the $Zb\bar{b}$ couplings: mixing of the zero mode fermions (after EWSB) with KK states of equal  electric but different SU(2)$_L\times$U(1) charges, and mixing among the gauge bosons.

We first estimate the contribution of the fermion mixing effect after EW breaking.
Since the b gains a mass after EWSB, in principle one can no longer use the zero mode wave functions.
Most of the time, these corrections are of order $(m_b R')^2$, and so can be safely ignored.
However, in the spinorial model one encounters order $(m_t R')^2$ corrections.
The leading effect of EWSB will be to twist the zero mode wave functions between the L and R multiplets of a bulk spinor via the Wilson line matrix.  The effect of this twisting (for simplicity, we set $\tilde{M}_u=\tilde{M}_d=0$) will be that the left handed bottom
quark will end up partly living in the down-type component of $\chi_{b_u^c}$.
The reason why there is a component in $\chi_{b_u^c}$ (but not in $\chi_{b_q^c}$ or $\chi_{b_d^c}$) is that the UV brane boundary
conditions only allow a non-vanishing component in this this mode.
This non-vanishing component in $\chi_{b_u^c}$ will give the leading correction to the $Zb\bar{b}$ vertex.
The coupling of the zero mode of $Z$ (in the limit it is flat) is given by $g_{Zf\bar{f}}\sim (T_{3L} -\sin^2\theta Q)$.
Since the electric charges of all components that mix are the same the only correction comes from the deviation in the coupling to $T_{3L}$, which in our case is due to the fact that the $\chi_{b_u^c}$ component couples to $T_{3R}$ and not to $T_{3L}$.
So the relative deviation can be estimated to be (where $s^2=1-c^2= \sin^2 \frac{v}{2f}$).
\begin{equation}
\frac{\delta g_{Z b_L \bar{b_L}}}{g_{Zb_L\bar{b_L}}} \sim \frac{\sin^2(\ti v/2f) \tilde{m}_u^2}{f_u^2\left[ \frac{1}{f_q^2} +\frac{\tilde{m}_u^2}{f_u^2}+\frac{\tilde{m}_d^2}{f_d^2}\right]}
\end{equation}
This expression can be related to the formula for the top mass, and so we find
\begin{equation}
\frac{\delta g_{Zb\bar{b}}}{g_{Zb\bar{b}}} \sim
\frac{(m_{top}R')^2}{f_u^2 f_{-u}^2} \sim \frac{(m_{top}R')^2}{(1-4
c_u^2)}
\end{equation}
For the range of interest $R'\sim 1-2$ TeV and $|c_u|<1/2$ we will find a correction that is always at least a percent, and cannot be removed.
For the other two models the effect does not occur % because $b_L$ has the same quantum numbers under $T_L^3$ and $T_R^3$.
due to  the custodial protection proposed in ref. \cite{ACDP}.

We move to the gauge contribution to the Zbb vertex.
That correction can be calculated from the formula
\beq
{\delta g_{Z b_L b_L} \over g_{Z b_L b_L}} =
m_Z^2  \int_{R}^{R'} (z/R)^{-2 c_q} \left (
- \int_R^z z' \log(z'/R)
+ {T_R^3 - {g^2\over g'{}^2} Y  \over T_L^3 - {g^2\over g'{}^2} Y} \left (\int_R^z z' \log(R'/R) \right )
\right )
\eeq
We know from the analysis of the Higgs potential that we need  $c_q < 1/2$.
For generic left and right quantum numbers of $b_L$, the result is of order $m_Z^2 R'{}^2 \log(R'/R)$, that is enhanced by the large logarithm.
This would lead to more stringent bounds than those from the S parameter.
However, in the models where $b_L$ is embedded in the bifundamental representation with $T_L^3 = T_R^3$ there is a cancellation of the large logs and the result is given by
\beq
{\delta g_{Z b_L b_L} \over g_{Z b_L b_L}} \approx
{m_Z^2 R'{}^2 \over 4} \left (1 - {4 \over (3- 2c_q)^2} \right )
\eeq
This is below the experimental sensitivity even for $R' = 1 \tev$.

%%%%%%%%%%%%%%%%%%%%%%%%%%%%%%%%%%%%%%%%%%%%%%%%%%%%%%%%%%%%%%%%%%%%%%%%%%%%%%%%%%
%%%%%%%%%%%%%%%%%%%%%%%%%%%%%%%%%%%%%%%%%%%%%%%%%%%%%%%%%%%%%%%%%%%%%%%%%%%%%%%%
\section{Flavor in GHU}
\label{sec:masses} \setcounter{equation}{0} \setcounter{footnote}{0}
%%%%%%%%%%%%%%%%%%%%%%%%%%%%%%%%%%%%%%%%%%%%%%%%%%%%%%%%%%%%%%%%%%%%%%%%%%%%%%%
%%%%%%%%%%%%%%%%%%%%%%%%%%%%%%%%%%%%%%%%%%%%%%%%%%%%%%%%%%%%%%%%%%%%%%%%%%%%%%%%%

We begin our study of the flavor structure of GHU models.
In this section we present a detailed discussion for the model of Section \ref{s.ff},
with each SM generation embedded in two fundamental and one adjoint SO(5) multiplet.
The other two models lead to a very similar flavor structure, and we will later comment on the differences.

%%%%%%%%%%%%%%%%%%%%%%%%%%%%%%%%%%%%%%%%%%%%%%%%%%%%%%%%%%%%%%%%%%%%%%%%%%%%%%%%
\subsection{Fermion masses and mixing}
%%%%%%%%%%%%%%%%%%%%%%%%%%%%%%%%%%%%%%%%%%%%%%%%%%%%%%%%%5

We start by discussing the zero modes before EW breaking produces their mass terms.
One difference with respect to the standard RS scenario is the  presence of the IR boundary mass terms.
These do not give masses to the zero modes, but imply that the zero modes are embedded in several bulk multiplets.
In particular, the zero mode quark doublets $q_L$ are embedded in all $\Psi_{q,u,d}$ in \eref{fac},
while the up-type quark singlets $u_R$ lives in $\Psi_{q,u}$.
On the other hand, the down-type quark singlets $d_R$ lives only in $\Psi_{d}$.
We write
\beq
q_q(x,z) \to \chi_{q_q}(z) q_L(x)
\qquad
q_q(x,z) \to \chi_{q_u}(z) q_L(x)
\qquad
q_d(x,z) \to \chi_{q_d}(z) q_L(x)
\eeq
\beq
u_u^c(x,z) \to \psi_{u_u^c}(z) u_R(x)
\qquad
u_q^c(x,z) \to \psi_{u_q^c}(z) u_R(x)
\qquad
d_r(x,z) \to \psi_{d_r}(z) d_R(x)
\eeq
Recall that we drop the generation index, and $q_L,u_R,d_R$ are understood to be three-vectors in the generation space.
The zero-mode profiles $\chi(z),\psi(z)$ are 3x3 matrices $\chi_q = {\rm diag} (\chi_{q_1},\chi_{q_2},\chi_{q_3})$ that
determine how much of the zero mode resides in each 5D fermion.
Solving the equations of motion and the boundary conditions we find the left-handed profiles:
\begin{eqnarray}
\chi_{q_q}(z) &=& \frac{1}{\sqrt{R'}}
\left( \frac{z}{R}\right)^{2}
\left( \frac{z}{R'}\right)^{-c_q} f_{q}
\nonumber \\
\chi_{q_u}(z) &=& \frac{1}{\sqrt{R'}}
\left(\frac{z}{R}\right)^{2}
\left(\frac{z}{R'}\right)^{-c_u}\tilde{m}_u^\dagger f_{q}
\nonumber \\
\chi_{q_d}(z) &=& \frac{1}{\sqrt{R'}}
\left( \frac{z}{R}\right)^{2}
\left( \frac{z}{R'}\right)^{-c_d} \tilde{m}_d^\dagger f_{q}
\label{e.LHzero}
\end{eqnarray}
where  $f_q = {\rm diag} (f(c_{q_1}),f(c_{q_2}),f(c_{q_3}))$ and
$f(c)$ were  defined in \eref{rsff}.
Similarly, the zero mode profiles for the right handed up-type fields are given by
\begin{eqnarray}
\psi_{u^c_u}(z) &=& \frac{1}{\sqrt{R'}}
\left(\frac{z}{R}\right)^{2}
\left(\frac{z}{R'}\right)^{c_u}f_{-u}
\nonumber \\
\psi_{u^c_q}(z) &=& -\frac{1}{\sqrt{R'}}
\left(\frac{z}{R}\right)^{2}
\left(\frac{z}{R'}\right)^{c_q} \tilde{M}_u f_{-u},
\end{eqnarray}
Finally, the down-type zero modes are contained only in the
adjoints:
\beq \psi_{d_r}(z) = \frac{1}{\sqrt{R'}}
\left(\frac{z}{R}\right)^{2}  \left(\frac{z}{R'}\right)^{c_d} f_{-d} \, .
\eeq
The overall normalization has
been chosen such that  we recover the usual  normalized zero modes
\erefn{zm} in the limit when the boundary masses are set to zero. In
this form, our profiles  closely resemble the corresponding formulae
in the standard RS set-up. However, in this basis, the kinetic terms
for the zero modes are not diagonal. This kinetic mixing is
parameterized by the Hermitian  3 by 3 matrices
\begin{eqnarray}
K_q&=& 1+f_q \tilde{m}_u f_u^{-2} \tilde{m}_u^\dagger f_q
+f_q\tilde{m}_d f_d^{-2} \tilde{m}_d^\dagger f_q,
\nonumber \\
K_u&=&
1+f_{-u} \tilde{M}_u^\dagger f_{-q}^{-2} \tilde{M}_u f_{-u},
\nonumber \\
K_d&=& 1,
\label{kinmix}
\end{eqnarray}
For example, for the doublet zero modes the kinetic term is given
by $i \bar{q}_L(x) K_q \pa \! \!\! /  q_L(x)$. The kinetic mixing is
inevitable in GHU models: since all the SM flavor mixing must
originate from non-diagonal terms in the boundary mass terms,  at
least the matrix $K_q$ must be non-diagonal. \emph{This is an
important difference with respect to the original RS set-up that
will introduce additional contributions to flavor-violating processes.}

In GHU, the fermion masses originate from the bulk  kinetic terms
$\bar{\Psi} i D^M \Gamma_M \Psi$, where $D_z\to \partial_z -i g_5
A_z^a T^a$ and  $T^a$ are the SO(5) generators appropriate for a
given representation. When $A_z$ acquires a vev it produces a mass
term connecting the quarks living in the same SO(5) multiplet. For
the fundamental representation, the Wilson line marries the two up
quarks in the  bifundamental to the singlet up quark:
\beq \left(
\frac{R}{z}\right)^4  \frac{g_* v}{2} \frac{\sqrt{2}z }{R'}
\ov u^c(u  - \ti u ) \eeq
while for the adjoint representation, it
couples the triplets to the bifundamental, for example
\beq \left(
\frac{R}{z}\right)^4  \frac{g_* v}{2} \frac{\sqrt{2}z }{R'}
(\ov d_l - \ov d_r) d_d \eeq

Plugging in the zero mode profiles we find for the mass matrix  (in
the basis were the kinetic terms are not diagonal)
\bea m_{u} &= &
\frac{g_* v }{2 \sqrt 2}  f_q (\tilde{m}_{u}
-\tilde{M}_{u}) f_{-{u}} \nn
m_{d} &= & \frac{g_* v}{2 \sqrt 2}  f_q \tilde{m}_{d} f_{-{d}} \eea
To obtain the
actual masses and mixing angles one needs to first diagonalize the
kinetic mixing terms and rescale the fields. We decompose $K_a=V_a
N_a V_a^\dagger$ for $a=q,u,d$, where $N$ is a positive diagonal
matrix and $V$ a unitary matrix, and we define the corresponding
Hermitian matrix $H_a =  V_a N_a^{-1/2} V_a^\dagger$. The Hermitian
rotation of the zero modes, $q_L \to H_q q_L$, $u_R \to H_u  u_R$,
brings the kinetic terms to the canonical form. The  mass matrices
rotate into \bea m_{u}^{SM} &=&\frac{g_*  v }{2 \sqrt 2}
%[V_q N_q^{-1/2} V^\dagger_q]
H_q f_q (\tilde{m}_{u} -\tilde{M}_{u}) f_{-{u}} H_ u %[V_{u} N_{u}^{-1/2} V_u^\dagger] ,
\nn
m_{d}^{SM} &=&\frac{g_* v}{2 \sqrt 2}
%[V_q N_q^{-1/2} V^\dagger_q]
H_q f_q \tilde{m}_{d}  f_{-{d}} . \label{MCHmass} \eea In the next
step we decompose, as usual, the up and down mass  matrices  as
$m_{u,d}^{SM}=U_{L\ u,d} m_{u,d} U_{R\ u,d}^\dagger$, and we perform
a unitary rotation of the zero mode quarks so as to diagonalize the
mass matrix, $d_{L,R} \to U_{L,R \ d} d_{L,R}$, $u_{L,R} \to U_{L,R
\ u} u_{L,R}$ .

The flavor structure following from \ref{MCHmass} is  very similar
to that of the ordinary RS model with anarchic flavor structure, cf.(\ref{RSmass}).
The main difference between (\ref{MCHmass}) and
(\ref{RSmass}) is the appearance of the extra Hermitian matrices
$H_a$ that originate from the kinetic mixing.
Note that, since $N_a$ the eigenvalues of the kinetic mixing
matrices (\ref{kinmix}), all entries of $N_a$ are very close to 1
except maybe for the third generation. For this reason, the
hierarchical structure of the mass matrix will turn out to be very
similar to that in RS.

The quark masses that follow from (\ref{MCHmass}) are  as usual,
approximately equal to the diagonal elements.
They can be estimated by
\bea m_u &\sim& \frac{g_* v}{2 \sqrt {2}}\frac{(\tilde{m}_u-\tilde{M}) f_q f_{-u}}
{\sqrt{\left( 1+\frac{f_q^2 \tilde{m}_u^2}{f_u^2}+\frac{f_q^2
\tilde{m}_d^2}{f_d^2} \right) \left( 1+\frac{f_{-u}^2
\tilde{M}^2}{f_{-q}^2} \right)}} \nn m_d &\sim &
\frac{g_* v }{2 \sqrt{2}}\frac{\tilde{m}_d f_q f_{-d}} {\sqrt{\left(
1+\frac{f_q^2 \tilde{m}_u^2}{f_u^2}+\frac{f_q^2 \tilde{m}_d^2}{f_d^2}
\right)}} \eea
where $\ti m_{u,d},\ti M$ here denotes the typical amplitude of the entries in the corresponding mass matrix.
As in the standard RS, the mass hierarchies are set by $ f_q, f_{-u},f_{-d}$ and $g_*/2$ plays the similar role to the Yukawa coupling $Y_*$.
The coupling $g_*$ is however related to the experimentally measured weak coupling, see \eref{wc}, and cannot be varied, unless we consider large UV brane kinetic terms.
Another difference is the dependence on the boundary masses that enters both the numerator and the denominator.
The latter  may be significantly different than 1 only when $f_{x} \sim 1$, which is the case for the third
generation. In that case $\ti m_{u,d},\ti M$ saturate: further increasing it does not increase the mass.

We turn to discussing the mixing angles and how are they affected by the kinetic mixing.
Consider the doublet mixing matrix $K_q$.
$f_q$ is  hierarchical, $f_{q_1} \ll f_{q_2} \ll 1$, $f_{q_3}\sim 1$,
while $f_{u,d}$'s are all of order one (since it is $f_{-u,d}$ that sets the mass hierarchy).
Thus, $K_q \sim \delta_{ij} + \ti m^2 f_{q_i} f_{q_j}$, where we denote $\ti m^2 = \ti m_u^2 + \ti m_d^2$.
%This follows from the structure of the kinetic mixing matrix.
It follows that $N_q \sim (1,1,1+ \ti m^2)$ and $(V_q)_{12} \sim \ti
m^2 f_{q_1} f_{q_2}$, $(V_q)_{13} \sim \ti m^2 f_{q_1}/ (1 + \ti
m^2)$, $(V_q)_{23} \sim \ti m^2 f_{q_2}/ (1 + \ti m^2)$. At the end
of the day, the left hierarchy is set by the matrix $f_q H_q$
(rather than $f_q$ as in  RS) whose diagonal elements
are  of the form \beq \label{e.dupa1}
% [V_q N_q^{-1/2} V_q^\dagger]
H_q f_q \sim
\left( \begin{array}{ccc} f_{q_1} &  & \\  & f_{q_2} &  \\
&  & f_{q_3} (1 +  f_{q_3}^2 \ti m^2)^{-1/2}
\end{array} \right),
\eeq The off-diagonal terms in the above matrix are irrelevant  for
the following discussion. We can see that the corrections coming
from the kinetic mixing do not change the hierarchy of $f_q$.
There are, of course, ${\cal O}(1)$ corrections in the actual
numerical values of the $f$'s that are required to reproduce the
mass and mixing hierarchies, but their orders of magnitude are  unchanged so that implementation of flavor hierarchies is
analogous as in RS.
The only parametric difference between $f_q$ and \eref{dupa1} is that the third eigenvalue gets
suppressed by $1/\ti m$ as soon as $\ti m$ is larger than 1. We keep
track of that effect because, as the study of the Higgs potential
shows, the interesting parameter space with successful electroweak
breaking and the large enough top quark mass extends to  $\ti m
\sim$ few. This parametric dependence feeds into the left rotation
that diagonalize the SM mass matrix, \beq\label{equ:ulUD} (U_{L
u,d})_{12} \sim {f_{q_1} \over f_{q_2}} \quad (U_{L u,d})_{13} \sim
{f_{q_1} \over f_{q_3}}  (1 + f_{q_3}^2 \ti m^2)^{1/2} \quad (U_{L
u,d})_{23} \sim {f_{q_2} \over f_{q_3}}  (1 + f_{q_3}^2  \ti
m^2)^{1/2} \eeq The consequence is that the relation between $f_q$
and the CKM angles is slightly modified, $(1 + \ti m^2)^{1/2}
f_{q_1} \sim \lambda^3$, $(1 + \ti m^2)^{1/2} f_{q_1} \sim
\lambda^2$, where $\lambda$ is the Cabibbo angle.

By the same token,
\beq
\label{e.dupa2}
%[V_u N_u^{-1/2} V_u^\dagger]
H_u f_{-u} \sim
\left( \begin{array}{ccc} f_{-u_1} &  & \\  & f_{-u_2} &  \\
&  & f_{-u_3} (1 + f_{-u_3}^2 \ti M^2)^{-1/2} \end{array} \right),
\eeq and the elements of the right rotation matrix for the up-type
quark are \beq (U_{R u})_{12} \sim {f_{-u_1}  \over f_{-u_2}} \quad
(U_{R u})_{13} \sim {f_{-u_1}  \over f_{-u_3}} (1 +  f_{-u_3}^2\ti
M^2)^{1/2} \quad (U_{R u})_{23} \sim {f_{-u_2}  \over f_{-u_3}} (1 +
f_{-u_3}^2\ti M^2)^{1/2} \eeq Finally, the elements of the right
rotation matrix for the down-type quark are not affected by the
kinetic mixing, \beq (U_{R d})_{12} \sim f_{-d_1}/f_{-d_2} \quad
(U_{R d})_{13} \sim f_{-d_1}/f_{-d_3} \quad (U_{R d})_{23} \sim
f_{-d_2}/f_{-d_3} \eeq

%%%%%%%%%%%%%%%%%%%%%%%%%%%%%%%%%%%%%%%%%%%%%%%%%%%%%
%%%%%%%%%%%%%%%%%%%%%%%%%%%%%%%%%%%%%%%%%%%%%%%%%%%%%%
\subsection{Flavor constraints}
%\label{sec:FCNC} \setcounter{equation}{0} \setcounter{footnote}{0}
%%%%%%%%%%%%%%%%%%%%%%%%%%%%%%%%%%%%%%%%%%%%%%%%%%%%%%
%%%%%%%%%%%%%%%%%%%%%%%%%%%%%%%%%%%%%%%%%%%%%%%%%%%%%%

We are ready to evaluate the flavor constraints in the GHU models that originate from a tree-level exchange of the KK gluons.
To this end, we need to compute the couplings of the SM down-type quarks to the lightest KK gluon,
\begin{equation}
\label{e.ghukkgc}
g_{L}^{ij} \bar{d}_{L}^i \gamma_\mu G_{\mu (1)}  d_{L}^j
+ g_{R}^{ij} \bar{d}_{R}^i \gamma_\mu G_{\mu (1)}d_{R}^j
\end{equation}
As before, we introduce the diagonal matrix
$g_{x} \approx  g_{s*}(  -\frac{1}{\log R'/R} +  \gamma(c_x)  f_x^2)$, $x = q,-u,-d$,
that approximate the couplings of quarks in each 5D multiplets to the lightest KK gluon.
The complication inherent to the GHU models is that the zero mode quarks are contained in several 5D multiplets.
This is already one source of off-diagonal couplings.
Furthermore, on top of the unitary rotation that diagonalizes the SM mass matrix, the off-diagonal terms are affected by the Hermitian rotation that diagonalizes the kinetic terms.
At the end of the day the couplings can be written as
\begin{eqnarray}
g_L^{ij} & \approx&  \left[U_{L \ d}^\dagger H_q
\left( g_q
+  f_q \tilde{m}_u  f_u^{-2} g_u \tilde{m}_u^\dagger f_q
+  f_q \tilde{m}_d  f_d^{-2} g_d \tilde{m}_d^\dagger f_q \right) H_q U_{L\ d}\right]_{ij}
\nn
g_R^{ij} &\approx&  \left[U_{R \ d}^\dagger g_{-d}  U_{R \ d}\right]_{ij}
\end{eqnarray}
Compared to the RS formula \erefn{rsg}, the doublet quark off-diagonal couplings receive many additional contributions.
In spite of that, the RS-GIM mechanism is still at work, in the sense that the off-diagonal terms are always multiplied by the hierarchical matrix $f_q$.
The flavor-changing couplings are proportional to
\bea
g_L^{ds} &\sim& g_{s*}  (1 + \ti m^2) f_{q_1} f_{q_2}
\nn
g_L^{db} &\sim&  g_{s*} {1 + \ti m^2 \over (1 + f_{q_3}^2 \ti m^2)^{1/2}} f_{q_1} f_{q_3}
\nn
g_L^{sb} &\sim&  g_{s*} {1 + \ti m^2 \over (1 + f_{q_3}^2 \ti m^2)^{1/2}}  f_{q_2} f_{q_3}
\eea
\beq
(g_{R,d})_{ij} \sim   g_{s*} f_{-d_i}  f_{-d_j}
\eeq

We can now estimate the size of the FCNC four-fermion operators relevant for the for the Kaon mixing.
The LL operator is, just like in RS, suppressed by the CKM matrix elements,
\beq
C_K^{1} \sim  \frac{1}{6 M_G^2}  g_{s*}^2 |V_{ts}^* V_{td}|^2
f_{q_3}^4  { (1 + \ti m^2)^2 \over (1 + f_{q_3}^2 \ti m^2)^2}
\eeq
This bound has almost the same form as in the RS case, except for the last factor that depends on $\ti m$.
In RS one usually takes $f_{q_3} \sim 0.3$ (that is $c_{q_3}$ close to one half) to further suppress the LL operator.
In the case at hand, using the approximate expression for the top mass, we can relate (for $f_{-u_3} \sim 1$)
\beq
f_{q_3}  \sqrt{1 + \ti m^2 \over  1 + f_{q_3}^2 \ti m^2}
\sim
{2 \sqrt{2} m_t \over g_* v} {\sqrt{1 + \ti m^2} \sqrt{1 + \ti M^2} \over \ti m_u - \ti M}
\eeq
Taking $\ti m_{u,d}\sim \ti M \sim 1$ we can rewrite our estimate as
\begin{eqnarray}
C_K^{1} \sim  \frac{1}{( 5 \cdot 10^4 \,{\rm TeV})^2} \left(\frac{3\, {\rm TeV} }{M_G}\right)^2 % \left(\frac{f_{q_3} }{0.3}\right)^4
\end{eqnarray}
which shows that a $3 \tev$ KK gluon  satisfies the bound listed in Table \ref{tab:all}.

On the other hand, the LR contribution is
\beq
C_K^{4} = - 3 \,C_K^{5} \sim  {1 \over M_G^2} {g_{s*}^2 \over g_*^2}\frac{8 m_d m_s}{v^2} \frac{1 + \ti m^2 }{\tilde m_d^2}
\eeq
where this time we related $f$'s to down-type quark masses via $f_{q_i} f_{-d_i} \sim 2 \sqrt{2} m_d^i/g_* v \ti m_d$.
Again, this estimate is very similar as in the RS case, with $Y_*$ replaced by $g_*/2$.
The new parametric difference is the factor $(1 + \ti m^2)/\tilde m_d^2$,  which  can be traced back to the kinetic mixing.
This factor may enhance (but never suppress) the coefficient, if $\ti m_d < 1$ or $\ti m_u > \ti m_d$.
In consequence (when boundary kinetic terms are ignored),
\beq
C_K^{4} \sim  \frac{1}{(1.5 \cdot 10^4\, {\rm TeV})^2} \left(\frac{3\, {\rm TeV} }{M_G}\right)^2
\eeq
The numbers refer to the case when the boundary kinetic terms at the TeV scale are small,
so that $g_* \sim g \log^{1/2} (R'/R) \sim 4$, $g_{s*} \sim g_s \log^{1/2} (R'/R) \sim 6$.
We can see that $\Lambda_{Im C^4_K}$ is a factor of $\sim 11$ too small for a $3 \tev$ KK gluon.
To satisfy the bound $\Lambda_{Im C^4_K} > 16 \cdot 10^4\, {\rm TeV}$ (see Table~\ref{tab:all}) would require a KK scale of about $33$ TeV.
\begin{figure}[tb]
\begin{center}
\includegraphics[width=8cm]{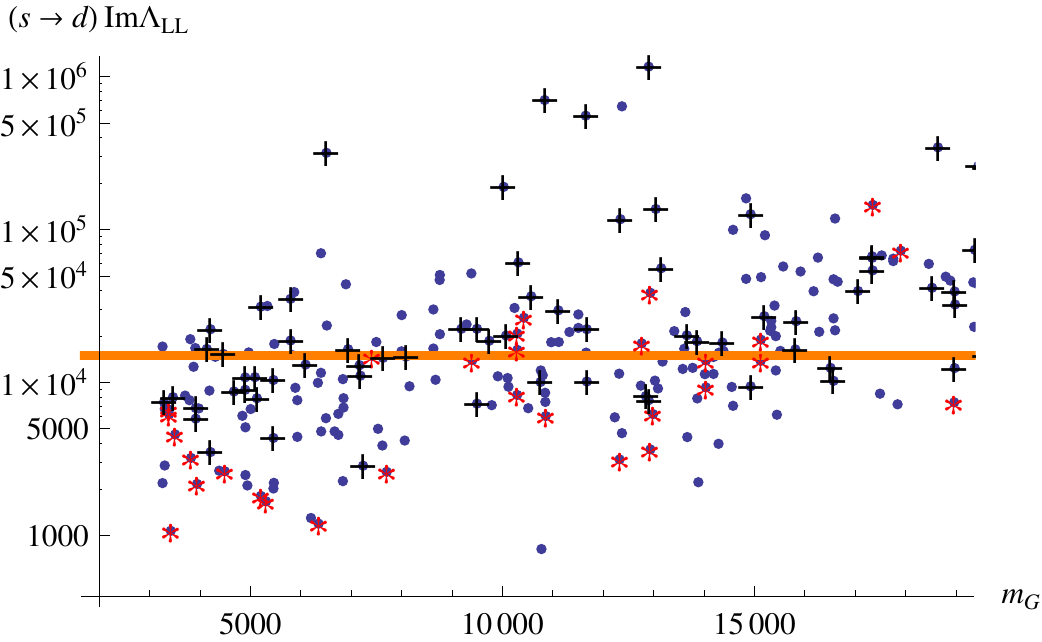}\quad
\includegraphics[width=8cm]{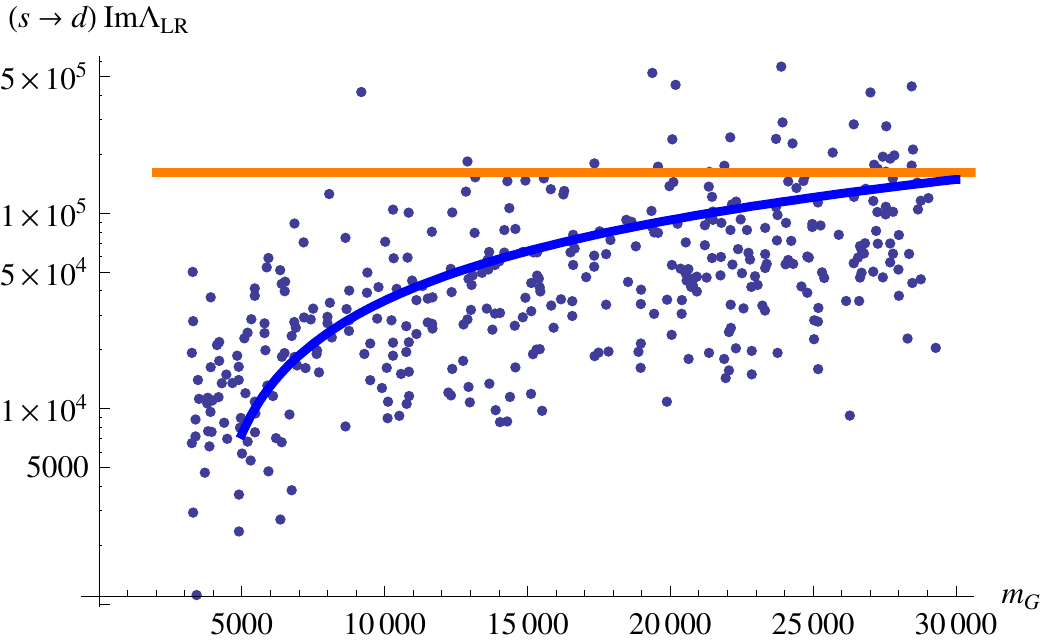}\quad
\caption{Scan of the effective suppression scale of ${\rm Im}C_{1K}$ (left panel)  ${\rm Im}C_{4K}$ (right panel) and  in the scenario with adjoint and fundamental bulk fields. In the left panel points with $\ti m_u>3.5$ are red (*) and points with $\ti m_u<1.5$ are black (+). All the points give the correct low-energy spectrum but most of the points with $m_G < 30$ TeV fail to satisfy the ${\rm Im}C_{4K}$ bound of $\Lambda>1.6 \cdot 10^5$ TeV. The blue line is a linear fit of the $M_G$ dependence.}  \label{fig:GHUbound}
\end{center}
\end{figure}
\begin{figure}[tb]
\begin{center}
\includegraphics[width=8cm]{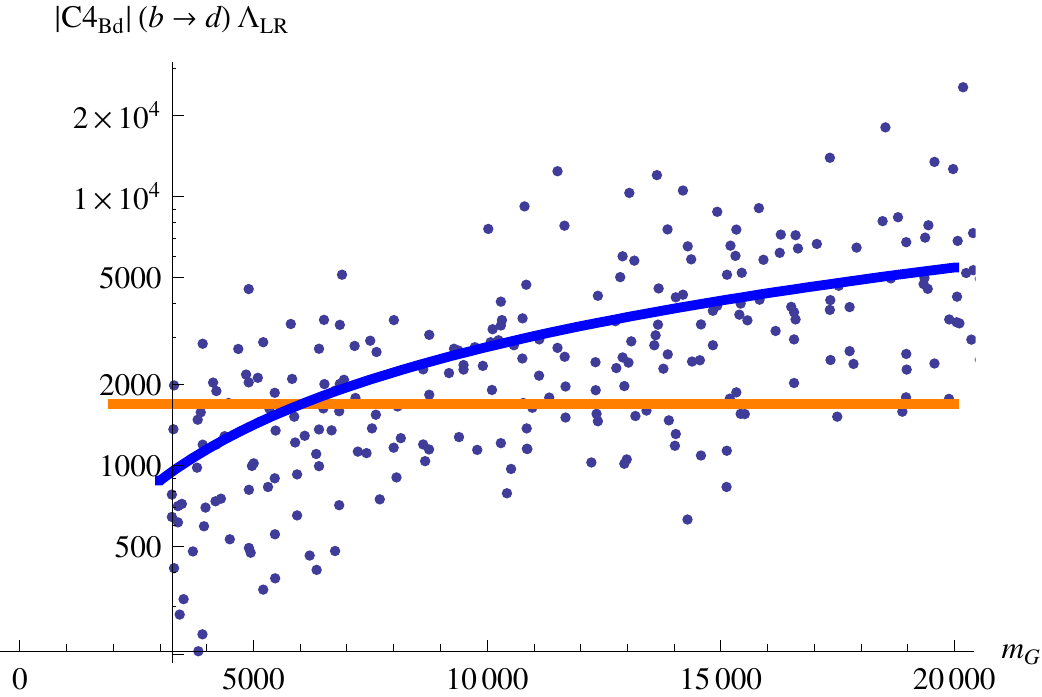}\quad
\includegraphics[width=8cm]{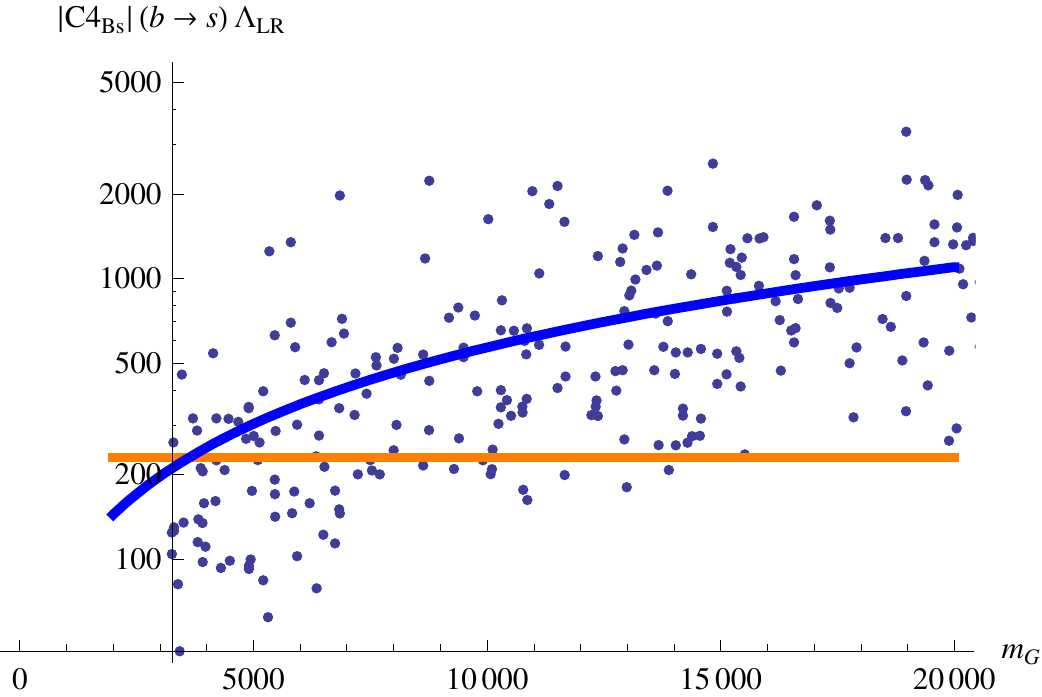}\quad
\caption{Scan of the effective suppression scale of $|C^4_{B_d}|$ (left panel)  $|C^4_{B_s}|$ (right panel) and in the scenario with adjoint and fundamental bulk fields. All the points give the correct low-energy spectrum and points with $m_G > 5$ TeV mostly satisfy the bounds. The blue lines are linear fits of the $M_G$ dependence.}  \label{fig:GHUsdbound}
\end{center}
\end{figure}

As in RS, we can change this estimate by playing with the boundary kinetic terms for $SU(2)_L$ and $SU(3)_C$.
Assuming a large bare UV coupling (small brane kinetic term) for $SU(3)_c$ and including the one-loop QCD running we can decrease $g_{s*}$, thus relaxing the bound by a factor of two.
On the other hand, we could make the coupling $g_*$ larger by adding a large brane kinetic term for $SU(2)_L$, that is to pick $r > 1$.
In the best of all worlds, assuming $g_{s*} \sim 3$ and $g_{*} \sim 4 \pi$ (the latter implies lack of  perturbative control over the Higgs sector)  we could lower the bound down to 5 TeV.
In a calculable framework it is not possible to lower the bound below 10 TeV.
These estimates demonstrate that the GHU framework with fully anarchic flavor is not a plausible scenario.

%%%%%%%%%%%%%%%%%%%%%%%%%%%%%%%%%%%%%%%%%%%%%%%%%%%%%%%%%%%%%%%%%%%%%%%%%%%%%
%%%%%%%%%%%%%%%%%%%%%%%%%%%%%%%%%%%%%%%%%%%%%%%%%%%%%%%%%%%%%%%%%%%%%%%%%%%%%
\subsection{Numerical scan}
\label{sec:numerics}
%%%%%%%%%%%%%%%%%%%%%%%%%%%%%%%%%%%%%%%%%%%%%%%%%%%%%%%%%%%%%%%%%%%%%%%%%%%%%
%%%%%%%%%%%%%%%%%%%%%%%%%%%%%%%%%%%%%%%%%%%%%%%%%%%%%%%%%%%%%%%%%%%%%%%%%%%%%
We have verified the above estimates of the flavor bounds by an
extensive numerical scan over the parameter space of the model. Our
first aim is to find realistic electro-weak symmetry breaking minima
of the effective potential. We are only interested in points with a
sizable gap between the EW scale and the scale of the KK modes
($v/f_\pi \lsim 0.3$). The parameters related by EWSB are: the bulk
masses of the third generation $c_{q_3}$,$c_{u_3}$,$c_{d_3}$, the
size of the brane masses $\ti m_u$, $\ti m_d$, $\ti M_u$, $\ti M_d$,
the KK scale $1/R'$. We keep the hierarchy $R/R'= 10^{-16}$ fixed.
Contrary to the RS case, we are not free to pick any combination,
since most choices do not lead to EWSB satisfying our criteria. As
we have discussed, $v/f_\pi$ is very sensitive to $q_{q_3},c_{u_3}$,
see e.g. fig.~\ref{fig:tunecu}. We therefore first randomly generate
a set of parameters with $c_{q_3} \in [0.2, 0.48]$, $1/R' \in [900,
12500]$ GeV, $c_{-d_3}= -0.55$, $|\ti m_u| \in [0.5, 5]$, $|\ti m_d|
\in [0.5, 2]$, and $r \in [0, 0.8]$. (For simplicity, we have set
$\ti M_u =\ti M_d = 0$). Knowing $R$,$R'$ and $r$, we determine
$v/f_\pi$ from the condition to reproduce $m_W$, see~(\ref{equ:mw}).
We then find a $c_{u_3}$ such that the minimum of the potential
really is at the value of $v/f_\pi$ specified before. Finally, we
want the theory to be calculable, which puts an upper bound on the
bulk $SO(5)$ coupling $g_{*}$. This can be equivalently expressed as
a bound on the number of colors of the dual CFT, see \eref{ncft},
and we require $N_{CFT} \gtap 5$.

The next step is to calculate the SM masses, mixing angles and KK FCNCs.
Given the size of $c_{q_3}$,$c_{u_3}$,$c_{d_3}$ and of $\ti m_u$ and
$\ti m_d$ we can determine the bulk masses of the first two generations.
First, $c_{q_1}$ and $c_{q_2}$ are fixed since left rotations (\ref{equ:ulUD}) need to have the same hierarchy
as the CKM. The remaining bulk masses $c_{u_{1,2}}$,$c_{d_{1,2}}$ can be fixed using (\ref{equ:crhRS}) requiring the mass eigenvalues to match the SM at TeV scales.

Now we randomly generate complex $3\times 3$ matrices $\ti m_u$,
$\ti m_d$ with eigenvalues approximately of the size as those used
in the calculation of the potential above. Using~(\ref{MCHmass}) we
then calculate the effective mass matrices $m_{u,d}^{SM}$. These
mass matrices only approximately reproduce the SM. In order to have
a completely realistic set of parameters we need to parameterize the
deviation from the SM and reject those $\ti m_{u,d}$ which deviate
too much. For this purpose we have defined
matrix norms that measure the distance between the generated and the physical values at 3 TeV. We
calculate the distance between
\begin{itemize}
\item The mass eigenvalues to the
the SM running masses at 3 TeV, see Table.~\ref{tab:quarkTeV}.
\item The moduli of the generated
CKM to
\begin{equation}
|V_{CKM}| \approx \left( \begin{array}{ccc} 0.97 &
0.226 & 0.0035 \\ 0.226 & 0.97 & 0.04
\\ 0.008 & 0.04 & 1 \end{array} \right),
\end{equation}
which we took from a recent tree level determination~\cite{Bona:2006sa}.
\item The amount of CP violation to the SM
Jarlskog invariant~\cite{J}
\begin{equation}
J  \approx 3 \cdot 10^{-5}
\end{equation}
\end{itemize}
 We finally accept only those random matrices $\ti m_{u,d}$ for which the maximal
  relative distance of any constraint is below $30\%$.

The results of the scan are displayed in Fig. \ref{fig:GHUbound}.
Most of the points with $m_G < 30$ TeV fail to satisfy the
$ImC_{4K}$ bound of $\Lambda>1.6 \cdot 10^5$ TeV, in agreement with
our analytical estimates. Just like in the RS case this bound is
somewhat dependent on the assumption that one makes on the boundary
kinetic terms for the gluon. For the two extreme cases the numerical
values for the supression scales are the following. In the  worst
case when the theory is barely perturbative, the bound is enhanced
to about 54 TeV, while in the best case, with no bare brane kinetic
terms to bound is reduced to about 17 TeV. We conclude that the
flavor constraints on the warped models with a pseudo-Goldstone
Higgs are even stronger that that in the standard RS with a TeV
brane localized Higgs.

In the remainder of this section we summarize the flavor
constraints  for the other two GHU models  considered in this paper.
The short summary is that in none of the models is the RS flavor bound relaxed, but rather strengthened.

%%%%%%%%%%%%%%%%%%%%%%%%%%%%%%%%%%%%%%%%%%%%%%%%%%%%%
%%%%%%%%%%%%%%%%%%%%%%%%%%%%%%%%%%%%%%%%%%%%%%%%%%%%%%
\subsection{Fermions in the spinorial}
\label{sec:twofives}
%\setcounter{equation}{0}
%\setcounter{footnote}{0}
%%%%%%%%%%%%%%%%%%%%%%%%%%%%%%%%%%%%%%%%%%%%%%%%%%%%%%
%%%%%%%%%%%%%%%%%%%%%%%%%%%%%%%%%%%%%%%%%%%%%%%%%%%%%%

For the model with each generation of the  SM fermions embedded in 3 SO(5) spinors the parameter space includes an additional boundary mass matrix $\ti M_d$ that introduces the kinetic mixing also to the  $d$-singlet quark sector.
The kinetic mixing matrices are given by
\begin{eqnarray}
K_q&=& 1+f_q \tilde{m}_u f_u^{-2} \tilde{m}_u^\dagger f_q +f_q
\tilde{m}_d f_d^{-2} \tilde{m}_d^\dagger f_q, \nonumber \\
K_u&=& 1+f_{-u} \tilde{M}_u^\dagger f_{-q}^{-2} \tilde{M}_u f_{-u}, \nonumber \\
K_d&=& 1+f_{-d} \tilde{M}_d^\dagger f_{-q}^{-2} \tilde{M}_d f_{-d},
\label{kinmixsp}
\end{eqnarray}
The SM up and down mass matrix take  the form
\begin{equation}
m_{u,d}^{SM}=\frac{g_* v }{4}
H_q f_q (\tilde{m}_{u,d} -\tilde{M}_{u,d}) f_{-{u,d}} H_{u,d}.
\label{MCHmasssp}
\end{equation}
There is an additional factor of $1/\sqrt{2}$ that is a group theoretical factor of the spinorial SO(5) representation.
The kinetic mixing induced by  $\ti M_d$ feeds into the right-rotation unitary matrix for the down quarks
\beq
(U_{R d})_{12} \sim {f_{-d_1}  \over f_{-d_2}}
\quad
(U_{R d})_{13} \sim {f_{-d_1}  \over f_{-d_3}} (1 +  f_{-d_3}^2\ti M^2)^{1/2}
\quad
(U_{R d})_{23} \sim {f_{-d_2}  \over f_{-d_3}} (1 +  f_{-d_3}^2\ti M^2)^{1/2}
\eeq
but the effect is negligible (unless $\ti M$ is very large) because $f_{-d_3} \ll 1$ to account for $m_b/m_t \ll 1$.
The only difference in flavor constraints is the consequence of the $1/\sqrt{2}$  factor in the SM mass matrix,
which makes the LR flavor bounds a factor of $\sqrt{2}$ more {\it stringent}.

%%%%%%%%%%%%%%%%%%%%%%%%%%%%%%%%%%%%%%%%%%%%%%%%%%%%%
%%%%%%%%%%%%%%%%%%%%%%%%%%%%%%%%%%%%%%%%%%%%%%%%%%%%%%
\subsection{Fermions in four fundamentals}
\label{sec:fourfives}
%\setcounter{equation}{0}
%\setcounter{footnote}{0}
%%%%%%%%%%%%%%%%%%%%%%%%%%%%%%%%%%%%%%%%%%%%%%%%%%%%%%
%%%%%%%%%%%%%%%%%%%%%%%%%%%%%%%%%%%%%%%%%%%%%%%%%%%%%%

The model with four fundamental  representation per generation
contains yet another matrix $\theta$ that is a source of flavor
violation. The kinetic mixing matrices are given by
\begin{eqnarray}
K_q&=& 1 + f_{q_u} (R'/R)^{c_{q_u}}\theta^\dagger  (R'/R)^{- c_{q_d}}f_{q_d}^{-2} (R'/R)^{- c_{q_d}} \theta (R'/R)^{c_{q_u}}f_{q_u}
+f_{q_u} \tilde{m}_u^\dagger f_{u}^{-2} \tilde{m}_u f_{q_u}
\nn &&
+f_{q_u} (R'/R)^{c_{q_u}} \theta^\dagger (R'/R)^{-c_{q_d}} \tilde{m}_d f_{d}^{-2} \tilde{m}_d^\dagger (R'/R)^{-c_{q_d}}\theta (R'/R)^{c_{q_u}} f_{q_u},
\nonumber \\
K_u&=& 1+f_{-u} \tilde{M}_u^\dagger f_{-q_u}^{-2} \tilde{M}_u f_{-u},
\nonumber \\
K_d&=& 1+f_{-d} \tilde{M}_d^\dagger f_{-q_d}^{-2} \tilde{M}_d f_{-d}, \label{kinmix45}
\end{eqnarray}
while the mass terms are \bea m_u^{SM} &=&  \frac{g_* v }{2 \sqrt 2}
H_{q} f_{q_u} (\tilde{m}_{u} -\tilde{M}_{u}) f_{-{u}} H_{u} \nn
m_d^{SM} &=&  \frac{g_* v }{2 \sqrt 2}  H_{q} f_{q_u}
(R'/R)^{c_{q_u}} \theta^\dagger (R'/R)^{- c_{q_d}} (\ti m_d - \ti
M_d) f_{-d} H_d \eea There are several new possible flavor effects
here. First of all, the mixing matrix $\theta$ shows, and it is
accompanied by the large factor   $ (R'/R)^{c_{q_u}}$ on the
outside. This factor perfectly counteracts the hierarchical
suppression generated by the matrix $f_q$. In consequence, as long
as  $\theta$ is an arbitrary anarchic matrix,  we definitely lose
the RS hierarchic structure of the fermion mass matrix. The origin
of this is easy to understand. Here we have two copies of the left
handed doublets, and we are removing one combination with an
additional right handed doublet on the UV brane. If the matrix
$\theta$ is anarchic, then  we introduce a large flavor violating
effect into the elementary sector, that is unsuppressed by the
mixing of elementary and composite states.

In order for the UV physics to maintain  an SU(3)$_Q$ flavor
symmetry for the doublets we need to assume that  $\theta$ is, to a
good approximation, proportional to the unit matrix. In the
following we set  $\theta \to \Theta 1_{3x3} $ in all expressions,
where $\Theta$ is a c-number. This is in fact a similar assumption
as the one implicitly makes in RS and the other two GHU models: the
elementary sector is flavor symmetric (that is there are no large
flavor violating kinetic mixing terms on the Planck brane), and only
the CFT gives rise to flavor violations. With this ansatz, the
flavor structure becomes very similar to that in the spinorial
model. There is still another new effect, however, which is the
presence of two sets of $c_q$. As a consequence, in the expression
for $m_{u,d}^{SM}$ the left hierarchy is not set by the same
functions: for the up-type masses it is set by $f_{c_{q_u}}$ while
for the down-type by $f_{c_{q_u}} (R'/R)^{c_{q_u}- c_{q_d}}$. This
effect can be used as an alternative way to explain the large
isosping breaking in the third generation. Instead of taking
$f_{-d_3} \ll 1$, that is assuming $b_R$ is mostly elementary, we
can choose $f_{-d_3} \sim 1$ and $c_{q_u^3} < c_{q_d^3}$ to obtain
$m_b/m_t \ll 1$. Unfortunately, this new avenue does not seem to
lead to suppressing dangerous four-fermion operators. The
coefficients $C_K^4$ can be estimated as \beq C_K^{4}  \sim
(R'/R)^{(c_{q_d^1}- c_{q_1})+ (c_{q_d^2}- c_{q_2})} {1 \over M_G^2 }
{g_{s*}^2 \over g_*^2 }\frac{8 m_d m_s}{v^2} \frac{1 + \ti m^2
}{\tilde m_d^2} \eeq where $c_{q_i} = \min (c_{q_u^i},c_{q_d^i})$.
For $c_{q_u^{1,2}} < c_{q_d^{1,2}}$ we are able to enhance the
coefficient $C_K^4$, but it's not exactly what we want.

%%%%%%%%%%%%%%%%%%%%%%%%%%%%%%%%%%%%%%%%%%%%%%%%%%%%%
%%%%%%%%%%%%%%%%%%%%%%%%%%%%%%%%%%%%%%%%%%%%%%%%%%%%%%
\section{Conclusion}
\label{sec:con}
\setcounter{equation}{0}
\setcounter{footnote}{0}
%%%%%%%%%%%%%%%%%%%%%%%%%%%%%%%%%%%%%%%%%%%%%%%%%%%%%%
%%%%%%%%%%%%%%%%%%%%%%%%%%%%%%%%%%%%%%%%%%%%%%%%%%%%%%

In this paper we studied flavor physics in the framework of 5D
warped GHU models that provide a  dual realization of a composite
pseudo-Goldstone Higgs.
This is an extension of the standard RS scenario,
that makes the electroweak symmetry breaking dynamical and fully calculable.

The flavor structure of GHU models turns out to be quite similar to
that in RS. The hierarchical structure of the quark masses and the
CKM matrix  appears as a consequence of different localization of
zero modes in the extra dimension. The RS-GIM mechanism is
operating, in the sense that the coefficients of effective
four-fermion $\Delta F = 2$ operators induced by the tree-level KK
mode exchange are suppressed by the small off-diagonal CKM matrix
elements (LL operators), or by the light quark masses (LR
operators).

The GHU models introduce, however, new contributions to flavor violation
that are of the same order of magnitude as those in RS. The reason
is that the larger set of local symmetries (that is crucial to
realize the Higgs as a pseudo-Goldstone boson) imply a different
realization of the zero-mode fermionic sector. In particular, the
zero modes must be embedded in more than one bulk multiplet.
We call it the kinetic
mixing because in the original flavor basis in 5D it shows up as
the generation mixing via the zero modes kinetic terms.
This effect is controlled by the IR boundary masses - the same Lagrangian
parameters that control also the mass matrix. The kinetic
mixing feeds into the effective four-fermion $\Delta F = 2$
operators, parametrically enhancing its coefficients by a factor of few.

Just like in RS, the strongest bound on the KK scale comes from the
imaginary  part of the LR $(sd)^2$ operator, that affects CP
violation in the kaon sector. We find that for anarchic boundary
mass matrices, the generic bounds on the lightest KK gluon mass is
of order $30 \tev$. Such a large KK scale of course undermines the
motivations for the GHU models.
This implies that GHU does not make sense without additional flavor symmetries. There are
several ways one could suppress the dangerous FCNC contributions
(apart from assuming accidental cancelations). For example in the
models of~\cite{ourflavor} one imposes a bulk flavor symmetry, and
all mixing originates from Planck brane localized kinetic mixing
terms. As a result one can construct models where all tree-level
FCNC's are absent (a genuine GIM mechanism), but of course in this
case one gives up on the explanation of the fermion mass hierarchy.
A recent proposal suggests to truncate the bulk at a scale of about $10^3$
TeV which softens the flavor problem by effectively  reducing $g_{s*}$~\cite{Davoudiasl:2008hx}. The obvious price to pay is giving up on an explanation for
the hierarchy between weak and Planck scale.
An intermediate solution could be to try to construct flavor models
that do explain both hierarchies, but have some partial flavor
symmetries left over~\cite{FPRa,inpreparation}. One possible ansatz
that would alleviate the constraints while still giving rise to the
hierarchies would be to assume that the boundary masses in the down
sector $\ti m_d$, $\ti M_d$ are proportional to the unit matrix,  so
that all the CKM mixing originate  from non-diagonal elements in
$\ti m_u$, $\ti M_u$. This suppresses the right rotation matrix
$U_{R\ d}$, so that the LR operator is also suppressed.

\section*{Acknowledgements}

We would like to thank Kaustubh Agashe, Roberto Contino, Sebastian
J\"ager, Gilad Perez,  Jose Santiago, Martin Schmaltz, Luca Silvestrini, and Carlos Wagner for useful discussions.
We are grateful to  Kaustubh Agashe, Roberto Contino and Gilad Perez for important comments on the manuscript.
The research of C.C. and A.W. is supported in part by the NSF grant
PHY-0355005 at Cornell and PHY05-51164 at the KITP. A.F. is
partially supported by the European Community Contract
MRTN-CT-2004-503369 for the years 2004--2008.

\appendix

\section*{Appendix}

%%%%%%%%%%%%%%%%%%%%%%%%%%%%%%%%%%%%%%%%%%%%%%%%%%%%%%%%%%%%%%%%%%%%%%%%%%%%%
\section{KK gluon sums}
\setcounter{equation}{0}
\label{a.kkgs}
%%%%%%%%%%%%%%%%%%%%%%%%%%%%%%%%%%%%%%%%%%%%%%%%%%%%%%%%%%%%%%%%%%%%%%%%%%%%%%
In this appendix we present analytical formulas that allow one to include the contribution of the entire KK gluon tower to the effective four-fermion operators.
We start with couplings of the $n$-th KK gluon to the quark eigenstates
\begin{equation}
\label{e.kkgc2}
g_{L,n}^{ij} \bar{q}_{L}^i \gamma_\mu G^{\mu (n)} q_{L}^j
+ g_{R,n}^{ij} \bar{q}_{R}^i \gamma_\mu G^{\mu (n)} q_{R}^j
\end{equation}
The couplings are given by
\beq
g_{L,n}^{ij} = g_{s*} R^{1/2}\int_R^{R'} f_n(z) h_{L}^{ij}(z)
\qquad
g_{R,n}^{ij} = g_{s*} R^{1/2}\int_R^{R'} f_n(z) h_{R}^{ij}(z)
\eeq
Above, $a(z)$ is the warp factor (that we keep arbitrary here), $f_n(z)$ is the profile of the n-th KK gluon, and $g_5$ is the dimensionful bulk coupling of $SU(3)$  color.
For vanishing brane kinetic terms, the SM strong coupling is given by $g_s^2 = g_{s*}^2 R/L$, where $L = \int_{R}^{R'} a(z)$.
The quark bilinear profiles $h_{L,R}^{ij}(z)$ are 3x3 matrices in the mass eigenstate basis.
They are related to the fermionic profiles:
\beq
h_{L}(z) = a^4(z) V_L^\dagger \sum_a \chi_{q_a}^\dagger(z) \chi_{q_a}(z) V_L
\qquad
h_{R}(z) = a^4(z) V_R^\dagger \sum_a \psi_{q_a}^\dagger(z) \psi_{q_a}(z)  V_R
\eeq
where the sum goes over all bulk multiplets in which the quark eigenstates is embedded,
and $V_{L,R}$ collectively denote all  Hermitian or unitary rotations that relate the original flavor basis to the mass eigenstate basis.
By orthogonality, the bilinear profiles satisfy $\int_R^{R'} h_{L,R}^{ij} = \delta_{ij}$.

The coefficients of the effective four-fermion operators are set by the sums of the form
\beq
\Sigma_{LR}^{ij,kl} = \sum_{n = 1}^\infty {g_{L,n}^{ij} g_{R,n}^{kl}  \over m_n^2}
\eeq
Strangely enough,  with the help of the methods of ref. \cite{HS} the sum can be evaluated for a general warp factor (even though there is no closed expression for particular $m_n$ and $g_n$).
The strategy is to 1) insert the integral expression for the couplings into the sum, 2) use the integrated equation of motion for $f_n(z)$, and  3) use the completeness relation for $f_n(z)$.
When the smoke clears, one is left with
\bea
\label{e.kkgsum}
\Sigma_{LR}^{ij,kl} = &
g_s^2  \left [
 L\int_R^{R'} a^{-1}(z) (\int_R^{z}  h_L^{ij}(z'))(\int_R^{z}  h_R^{kl}(z'))
\bnl
+\delta_{ij} \int_R^{R'}  h_R^{kl}(z) \int_R^z a^{-1}(z') \int_R^{z'} a(z'')
+  \delta_{kl}  \int_R^{R'} h_L^{ij}(z) \int_R^z a^{-1}(z') \int_R^{z'} a(z'')  \right ]
\nl
-  g_s^2  \delta^{ij} \delta^{kl} \left [
 \int_R^{R'} a^{-1}(z') \int_R^{z'} a(z'')  +   L^{-1} \int_R^{R'} a(z)  \int_R^z a^{-1}(z') \int_R^{z'} a(z'')
\right ]
\eea
The result is quite complicated,
but it simplifies considerably for flavor changing sums, which are of primary interest here:
\beq
\label{e.kkgsum2}
\Sigma_{LR}^{ij,kl} =
g_s^2  L  \int_R^{R'} a^{-1}(z) \left (\int_z^{R'} h_L^{ij}(z')\right)\left (\int_z^{R'}  h_R^{kl}(z')\right)
\quad i \neq j \quad k \neq l
\eeq
Thus, the whole sum is expressed by simple integrals of the warp factor and the fermion profiles.
In particular, for AdS geometry we take $a(z) = R/z$, and the fermionic profiles from \eref{zm}.
The AdS warp factor is sharply peaked towards IR. Therefore the integrals depend mainly on the IR value of the fermionic profiles, that also sets the fermion masses and mixing angles. This is the origin of the RS-GIM mechanism.

The coefficients of the operators relevant for the kaon sector are given by
\bea
C_{1K} &=& {1 \over 6}  g_s^2  \log(R'/R) \int_R^{R'} z \left (\int_z^{R'} h_L^{sd}(z')\right)\left (\int_z^{R'}  h_L^{sd}(z')\right)
\nn
C_{4K} &=& -  g_s^2  \log(R'/R) \int_R^{R'} z \left (\int_z^{R'}  h_L^{sd}(z')\right)\left (\int_z^{R'} h_R^{sd}(z')\right)
\nn
C_{5K} &=& -{1 \over 3} C_{4K}
\eea

In RS, the bilinear profiles of the down-type quarks are given by
\bea
h_{L,d}(z) &=&
R'{}^{-1} U_{L\ d}^\dagger  \left (\frac{R'}{z}\right)^{2 c_q} f_q^2  U_{L\ d}
%(V_L^d)^\dagger N_q a^{-1}(y)  e^{- 2 M_q y} N_q  V_L^d
\nn
h_{R,d}(z) &=& R'{}^{-1} U_{R\ d}^\dagger  \left (\frac{R'}{z}\right)^{-2 c_d} f_{-d}^2  U_{R\ d}
%(V_R^d)^\dagger N_d a^{-1}(y) e^{2 M_d y}   N_d V_R^d
%\nn
%h_L^{u}(y) &=&  (V_L^u)^\dagger N_q a^{-1}(y)  e^{- 2 M_q y} N_q V_L^u
% \nn  h_R^{u}(y) &=& (V_R^u)^\dagger N_u a^{-1}(y) e^{2 M_u y}   N_u V_R^u
\eea
Here, the non-diagonal elements are only due to the left and right unitary rotations that diagonalize the SM mass matrix.

For the GHU model with two fundamentals and an adjoint, the right-handed down-quark bilinear profile remains the same,
but the left-handed one becomes more complicated due to the kinetic mixing:
\beq
h_{L,d}(z) =
R'{}^{-1} U_{L\ d}^\dagger  H_q   \left [
 \left (\frac{R'}{z} \right)^{2 c_q} f_q^2
 +  f_q \ti m_u \left ( \frac{R'}{z}\right)^{2 c_u}   \ti m_u^\dagger f_q
 +  f_q \ti m_d  \left (\frac{R'}{z}\right)^{2 c_d}      \ti m_d^\dagger f_q
 \right ] H_q  U_{L\ d} .
\eeq
This time there are new sources of off-diagonal terms: the boundary masses $\ti m$ and the Hermitian rotations $H$.
%From the definition of the Hermitian rotation,  $\int_R^{R'} a^4 h_{L,d}^{ij} = \delta_{ij}$, as it should.

%%%%%%%%%%%%%%%%%%%%%%%%%%%%%%%%%%%%%%%%%%%%%%%%%%%%%%%%%%%%%%%%%%%%%%%%%%%%%
\section{SO(5) generators}
\setcounter{equation}{0}
\label{a.g}
%%%%%%%%%%%%%%%%%%%%%%%%%%%%%%%%%%%%%%%%%%%%%%%%%%%%%%%%%%%%%%%%%%%%%%%%%%%%%%

%&&&&&&&&&&&&&&&&&&&&&&&&&&&&&&&&&&&
\subsection{Spinorial}
%&&&&&&&&&&&&&&&&&&&&&&&&&&&&&&&&&&&
The smallest $SO(5)$ representation is the {\bf 4} spinorial.
The generators in a convenient basis:
\beq T_L^a = {1 \over 2} \left [ \ba{cc} \sigma^a & 0
\\
0 &  0 \ea \right ] \qquad T_R^a = {1 \over 2} \left [ \ba{cc} 0 & 0
\\
0 &  \sigma^a \ea \right ] \eeq \beq T_C^a = {i \over 2 \sqrt{2}}
\left [ \ba{cc} 0 & \sigma^a
\\
-  \sigma^a  &  0 \ea \right ] \qquad T_C^4 = {1 \over 2 \sqrt{2}}
\left [ \ba{cc} 0 & 1
\\
1 &  0 \ea \right ] \eeq
$T_L^a$ and $T_R^a$ generate the $SO(4) \equiv SU(2)_L \times SU(2)_R$ subgroup of
$SO(5)$ and $T_C^\ha$ are the coset generators.
The $T_3$ generators of the $SU(2)_L \times SU(2)_R$ subgroup are diagonal in this basis.
This makes transparent how the $SU(2)_L \times SU(2)_R$ quantum numbers are embedded in $\bf 4$:
\beq \Psi =
\left [
\ba{c} q_{+0}
\\
q_{-0}
\\
q_{0+}
\\
q_{0-} \ea  \right ] \qquad {\bf 4 = (2,1)\oplus(1,2) }
\eeq
where $\pm$ stands for $\pm 1/2$.
The Wilson-line  exponential $ e^{i \sqrt {2} h T_C^4}$ rotates
$q_{\pm 0}$ into $q_{0 \pm}$ and vice-versa:
\bea
q_{\pm 0} &\to&  \cos(h/2) q_{\pm 0} +   i \sin(h/2) q_{ 0\pm }
\nn
q_{0\pm}  &\to&  i\sin(h/2) q_{0 \pm} +  \cos(h/2) q_{\pm 0}
\eea

%&&&&&&&&&&&&&&&&&&&&&&&&&&&&&&&&&&&
\subsection{Fundamental}
%&&&&&&&&&&&&&&&&&&&&&&&&&&&&&&&&&&&
The 10 generators of the {\bf fundamental representation} in  an inconvenient basis:
\bea
T_{L,ij}^a & = & -{i \over 2} \left [ {1
\over 2} \eps^{abc}(\delta_i^b \delta_j^c - \delta_j^b \delta_i^c) +
(\delta_i^a \delta_j^4 - \delta_j^a \delta_i^4) \right]  \quad  a =
1 \dots 3 \nn T_{R,ij}^a & = & -{i \over 2} \left [ {1 \over 2}
\eps^{abc}(\delta_i^b \delta_j^c - \delta_j^b \delta_i^c)
-(\delta_i^a \delta_j^4 - \delta_j^a \delta_i^4)
\right]  \quad  a = 1 \dots 3
\nn T_{C,ij}^\ha & = & -{i \over \sqrt 2} \left [ \delta_i^\ha
\delta_j^5 - \delta_j^\ha \delta_i^5)
\right]  \qquad  \qquad  \ha = 1 \dots 4
\eea
The generators are normalized as $\tr T^\alpha T^\beta = \delta^{\alpha\beta}$.
The $T_3$ generators of $SU(2)_L \times SU(2)_R$ are non-diagonal in this basis.
The quantum numbers of the 5-vector components can be found as:
\beq Q =
{1 \over \rt} \bvec q_{++} +
q_{--}
\\
i q_{++} -  i q_{--}
\\
q_{+-} + q_{-+}
\\
i q_{+-} -  i q_{-+}
\\
\rt q_{00} \evec
\qquad  {\bf 5 = (2,2)\oplus(1,1) }
\eeq
The Wilson-line transformation $\exp( i \sqrt{2} h T_C^4)$  rotates $(q_{+-}, q_{-+},q_{00})$ into one another:
\bea
q_{+-} &\to&   { 1 +  \cos(h) \over 2} q_{+-} +   {1 -  \cos(h) \over 2} q_{-+}  - i {\sin(h)  \over \sqrt 2} q_{00}
\nn
q_{-+} &\to&   {1 -  \cos(h) \over 2} q_{+-} +   {1 +  \cos(h) \over 2}q_{-+}  +  i {\sin(h) \over \sqrt 2} q_{00}
\nn
q_{00} &\to&   -  i {\sin(h)  \over \sqrt 2} q_{+-}  +   i{ \sin(h) \over \sqrt 2} q_{-+}  +  \cos(h)q_{00}
\eea

%&&&&&&&&&&&&&&&&&&&&&&&&&&&&&&&&&&&
\subsection{Adjoint}
%&&&&&&&&&&&&&&&&&&&&&&&&&&&&&&&&&&&

The antisymmetric tensor $\Phi$ transforms as $\Phi \to \Omega \Phi \Omega^T$.
We can represent the tensor as: \beq \Phi = \Phi_L^a T_L^a + \Phi_R^a T_R^a + \Phi_C^\ha T_c^\ha \eeq
where $T$ are $SO(5)$ generators in the fundamental representation.
The SO(5) commutation relations
\beq \label{e.so5cr} [T_L^a,T_L^b ] = i \eps^{abc} T_L^c \qquad
[T_R^a,T_R^b ] = i \eps^{abc} T_R^c \qquad [T_L^a,T_R^b ] = 0 \eeq
\beq [T_C^a,T_C^b ] = {i \over 2} \eps^{abc} (T_L^c + T_R^c) \qquad
[T_C^a,T_C^4 ] = {i  \over 2} (T_L^a - T_R^a) \eeq \beq
[T_{L,R}^a,T_C^b ] = {i \over 2} \left ( \eps^{abc} T_C^c \pm
\delta^{ab} T_C^4 \right ) \qquad [T_{L,R}^a,T_C^4 ] = \mp {i  \over
2} T_C^a \eeq
imply that $\Phi_L$ is an $SU(2)_L$ triplet, $\Phi_R$ is an $SU(2)_R$ triplet and $\Phi_C$ is an $SU(2)_L \times SU(2)_R$ bifundamental, and the embedding of the  quantum numbers is
\beq {\bf (3,1):} \quad \Phi_L^1 \pm i
\Phi_L^2 \to (\pm 1,0) \quad \Phi_L^3 \to (0,0) \eeq \beq {\bf
(1,3):} \quad \Phi_R^1 \pm i \Phi_R^2 \to (0,\pm 1) \quad \Phi_R^3
\to (0,0) \eeq \beq {\bf (2,2):} \quad \Phi_C^1 \pm i \Phi_C^2 \to
(\pm 1/2,\pm 1/2) \quad \Phi_C^3 \pm i \Phi_C^4  \to (\pm 1/2, \mp
1/2) \eeq \beq {\bf 10} = {\bf (3,1)} \oplus {\bf (1,3)} \oplus
{\bf(2,2)}
\eeq
The Wilson-line transformation rotates the axial combination of the left and right triplets into the bifundamental
\bea
{\Phi_L^a + \Phi_R^a \over \sqrt 2} &\to&  {\Phi_L^a + \Phi_R^a \over \sqrt 2}
\nn
{\Phi_L^a - \Phi_R^a \over \sqrt 2} &\to&  {\Phi_L^a - \Phi_R^a \over \sqrt 2} \cos (h)  +  \Phi_C^a \sin(h)
\nn
\Phi_C^a &\to&  - {\Phi_L^a - \Phi_R^a \over \sqrt 2} \sin (h) + \Phi_C^a \cos(h)
\eea

%&&&&&&&&&&&&&&&&&&&&&&&&&&&&&&&&&&&66

\end{document}